\documentclass[useAMS,usenatbib]{mn2e}

\def\asca{{\it ASCA\/}}

\def\chandra{{\it Chandra\/}}

\def\hst{{\it {\it HST}\/}}

\def\rosat{{\it ROSAT\/}}

\def\spitzer{{\it Spitzer\/}}

\def\xmm{{\it XMM-Newton\/}}

\def\xray{\hbox{X-ray}}
\def\ecdfs{\hbox{E-CDF-S}}
\def\cdfs{\hbox{CDF-S}}
\def\cdfn{\hbox{CDF-N}}
\def\aegisx{\hbox{AEGIS-X}}
\def\etal{{et\,al.}}
\def\ae{{\ttfamily AE}}

\def\ltsima{$\; \buildrel < \over \sim \;$}
\def\simlt{\lower.5ex\hbox{\ltsima}}
\def\gtsima{$\; \buildrel > \over \sim \;$}
\def\simgt{\lower.5ex\hbox{\gtsima}}
\def\kms{\ifmmode{~{\rm km~s^{-1}}}\else{~km s$^{-1}$}\fi}
\def\lsim{\lower0.3em\hbox{$\,\buildrel <\over\sim\,$}}
\def\gsim{\lower0.3em\hbox{$\,\buildrel >\over\sim\,$}}

\def\msol{$M_\odot$}
\def\h2{H$_2$}
\def\flux{ergs~cm$^{-2}$~s$^{-1}$}
\def\xlum{ergs~s$^{-1}$}

\def\urltilda{\kern -.15em\lower .7ex\hbox{\~{}}\kern .04em}

\def\Lx{$L_{\rm X}$}
\def\aap{A\&A}
\def\apj{ApJ}
\def\apjl{ApJL}
\def\apjs{ApJS}
\def\aj{AJ}
\def\mnras{MNRAS}
\def\araa{ARA\&A}
\def\pasj{PASJ}
\def\nat{Nature}

\title[Chandra Catalogs for SSA22]{The {\itshape Chandra} Deep Protocluster Survey: Point-Source Catalogs for a 400~ks Observation of the {\itshape z} = 3.09 Protocluster in SSA22.} 
\author[LEHMER ET AL.]
{B.~D.~Lehmer,$^1$
D.~M.~Alexander,$^1$
S.~C.~Chapman,$^2$
Ian~Smail,$^3$
F.~E.~Bauer,$^4$
\newauthor
W.~N.~Brandt,$^5$
J.~E.~Geach,$^1$
Y.~Matsuda,$^1$
J.~R.~Mullaney,$^1$
\& A.~M.~Swinbank$^4$ \\
$^1${Department of Physics, University of Durham, South Road, Durham, DH1 3LE, UK}\\
$^2${Institute of Astronomy, Madingley Road, Cambridge CB3 0HA, UK}\\
$^3${Institute for Computational Cosmology, Department of Physics, Durham University, South Road, Durham DH1 3LE, UK.}\\
$^4${Columbia Astrophysics Laboratory, Columbia University, Pupin Labortories, 550 W. 120th St., Rm 1418, New York, NY 10027, USA}\\
$^5${Department of Astronomy \& Astrophysics, 525 Davey Lab, The Pennsylvania State University, University Park, PA 16802, USA}\\
}

\date{}
\pagerange{\pageref{firstpage}--\pageref{lastpage}} \pubyear{2009}

\usepackage{times}
\usepackage{graphicx}

\begin{document}

\label{firstpage}

\maketitle

%
\begin{abstract}
%

We present \xray\ point-source catalogs for a deep $\approx$400~ks \chandra\
\hbox{ACIS-I} exposure of the SSA22 field.  The observations are centred on a
$z = 3.09$ protocluster, which is populated by Lyman break galaxies (LBGs),
Ly$\alpha$ emitters (LAEs), and extended Ly$\alpha$-emitting blobs (LABs).  The
survey reaches ultimate (\hbox{3~count}) sensitivity limits of
$\approx$5.7~$\times 10^{-17}$~\flux\ and $\approx$3.0~$\times 10^{-16}$~\flux\
for the \hbox{0.5--2~keV} and \hbox{2--8~keV} bands, respectively
(corresponding to $L_{\rm 2-10~keV} \approx 5.7 \times 10^{42}$ and
\hbox{$L_{\rm 10-30~keV} \approx 2.0 \times 10^{43}$~\xlum} at $z = 3.09$,
respectively, for an assumed photon index of $\Gamma = 1.4$).  These limits
make SSA22 the fourth deepest extragalactic \chandra\ survey yet conducted, and
the only one focused on a known high redshift structure.  In total, we detect
297 \hbox{X-ray} point sources and identify one obvious bright extended \xray\
source over a $\approx$330~arcmin$^2$ region.  In addition to our \xray\
catalogs, we provide all available optical spectroscopic redshifts and
near-infrared and mid-infrared photometry available for our sources.  The basic
\hbox{X-ray} and infrared properties of our \chandra\ sources indicate a
variety of source types, although absorbed active galactic nuclei (AGNs) appear
to dominate.  In total, we have identified 12 \xray\ sources (either via
optical spectroscopic redshifts or LAE selection) at \hbox{$z =$~3.06--3.12}
that are likely to be associated with the SSA22 protocluster.  These sources
have \xray\ and multiwavelength properties that suggest they are powered by AGN
with \hbox{0.5--8~keV} luminosities in the range of
\hbox{$\approx$$10^{43}$--$10^{45}$}~\xlum.  We have analysed the AGN fraction
of sources in the protocluster as a function of local LAE source density and
find suggestive evidence for a correlation between AGN fraction and local LAE
source density (at the $\approx$96~per~cent confidence level), implying that
supermassive black hole growth at $z \approx 3$ is strongest in the highest
density regions. 

%
\end{abstract}
%


%
\section{Introduction}
%

Deep \xray\ surveys with \chandra\ and \xmm\ have provided a unique perspective
on the cosmic evolution of accreting supermassive black holes (SMBHs) and
high-energy activity from normal galaxies (e.g., star-formation processes,
evolving stellar binaries, and hot gas cooling) over significant fractions of
cosmic history (see Fig.~1 and Brandt \& Hasinger 2005 for a review).  Due to
limitations originating from point-source confusion for \xmm\ at
\hbox{0.5--2~keV} flux levels below $\approx$$10^{-16}$~\flux, the deepest
views of the extragalactic \xray\ Universe have come exclusively from
deep \chandra\ surveys with exposures \hbox{$\simgt$200~ks}.  

The deepest \chandra\ surveys to date are the $\approx$2~Ms \chandra\ Deep
Field-North (\cdfn; Alexander \etal\ 2003) and $\approx$2~Ms \chandra\ Deep
Field-South (\cdfs; Luo \etal\ 2008) surveys (hereafter, the \chandra\ Deep
Fields; CDFs).  These have been supplemented by additional deep \chandra\
surveys (e.g., the $\approx$250~ks Extended \cdfs\ [\ecdfs; Lehmer \etal\ 2005]
and the 0.2--0.8~Ms Extended Groth Strip [\aegisx; Laird \etal\ 2009]), which
cover larger solid angles than the CDFs and provide important \xray\
information on rarer source populations.  These deep \chandra\ surveys have
been chosen to lie at high Galactic latitudes in regions of the sky that have
relatively low Galactic columns ($\simlt$10$^{20}$~cm$^{-2}$) and few local
sources in the field, as to gain a relatively unobscured and unbiased
perspective of the distant ($z \simgt 0.1$) Universe.  Despite these growing
resources, there has not yet been an equivalent survey targeting very
high-density regions of the $z \simgt 2$ Universe where the most massive SMBHs
and galaxies are thought to be undergoing very rapid growth (e.g.,
Kauffmann~1996; De~Lucia \etal\ 2006).  In order to address this limitation and
study the growth of SMBHs and galaxies as a function of environment, we have
carried out a deep $\approx$400~ks \chandra\ survey (PI:~D.~M.~Alexander) of
the highest density region in the $z = 3.09$ SSA22 protocluster.

%
%
\begin{figure}
\centerline{
\includegraphics[width=9cm]{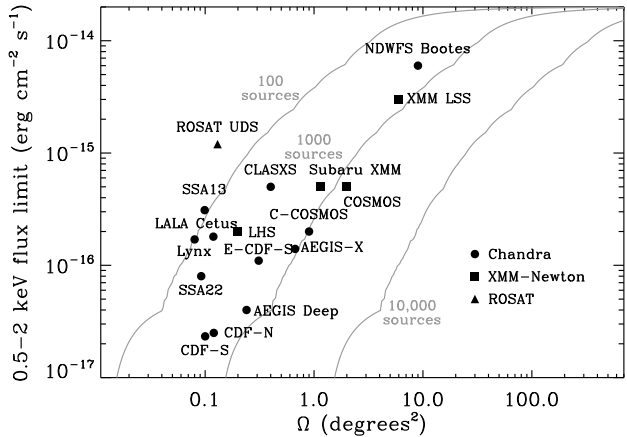}
}
\caption{
Locations of some well-known extragalactic \xray\ surveys conducted with
\chandra\ ({\it circles\/}), \xmm\ ({\it squares\/}), and \rosat\
({\it triangles\/}) in the \hbox{0.5--2~keV} flux limit versus solid-angle
$\Omega$ plane.  Each survey has a range of flux limits across its solid angle;
we have generally shown the most sensitive flux limit.  This plot has been
adapted from Fig.~1 of Brandt \& Hasinger (2005) to show the part of parameter
space most relevant for SSA22; see Table~1 of Brandt \& Hasinger (2005) for
references and descriptions of many of the surveys plotted here.  The gray
curves show contours regions of the flux limit versus solid angle plane where
$\approx$100, 1000, and 10,000 sources are expected to be detected (based on
the Bauer \etal\ 2004 number counts relation).
}
\end{figure}

The $z = 3.09$ SSA22 protocluster was originally identified by Steidel \etal\
(1998), who found a significant overdensity ($\approx$4--6 times higher surface
density than the field) within an $\approx$$8\times8$~arcmin$^2$ region
($\approx$$15\times 15$ comoving Mpc$^2$ at $z=3.09$) through spectroscopic
follow-up observations of $z \sim 3$ candidate Lyman break galaxies (LBGs).
Theoretical modelling indicates that the protocluster will evolve into a rich
cluster with a total mass $\simgt$$10^{15}$~\msol\ at the present day (e.g.,
Coma; see Steidel \etal\ 1998 for details).  Since its discovery, the
protocluster has also been found to contain a factor of $\approx$6 overdensity
of Ly$\alpha$ emitters (LAEs; Steidel \etal\ 2000; Hayashino \etal\ 2004;
Matsuda \etal\ 2005) and many remarkable bright extended Ly$\alpha$-emitting
blobs (LABs; Steidel \etal\ 2000; Matsuda \etal\ 2004), which are believed to
be sites of massive galaxy formation powered by starburst/AGN outflows (e.g.,
Bower \etal\ 2004; Geach \etal\ 2005, 2009; Wilman \etal\ 2005).  Therefore,
SSA22 is an ideal field for studying how SMBH growth depends on environment in
the $z \gg$~1 Universe.

In addition to the LBG and LAE surveys, SSA22 has become a premier
multiwavelength survey field.  The region has been imaged from space by \hst\
in a sparse mosaic of 12 ACS pointings using the F814W filter and by \spitzer\
over the entire field in four IRAC bands (3.6, 4.5, 5.8, and 8.0$\mu$m) and at
24$\mu$m by MIPS (see Fig.~2).  Additionally, the ground-based observations are
extensive and include imaging at radio (1.4~GHz from the VLA; Chapman \etal\
2004), submm (SCUBA 850$\mu$m, AzTEC 1.1mm, and LABOCA 870$\mu$m; Chapman
\etal\ 2001; Geach \etal\ 2005; Tamura \etal\ 2009), near-infrared ($J$, $H$,
and $K$ bands from UKIRT and Subaru; Lawrence \etal\ 2007; Uchimoto \etal\
2008), and optical wavelengths (e.g., Subaru $B$, $V$, $R$, $i^\prime$,
$z^\prime$, and NB497 bands; Hayashino \etal\ 2004).  A variety of
spectroscopic campaigns have been conducted (e.g., at Kitt Peak, Subaru, Keck,
and the VLT; Steidel \etal\ 2003; Matsuda \etal\ 2005; Garilli \etal\ 2008;
Chapman \etal\ 2009, in-preparation) and others are currently underway.

%
%

\begin{figure}
\centerline{
\includegraphics[width=9cm]{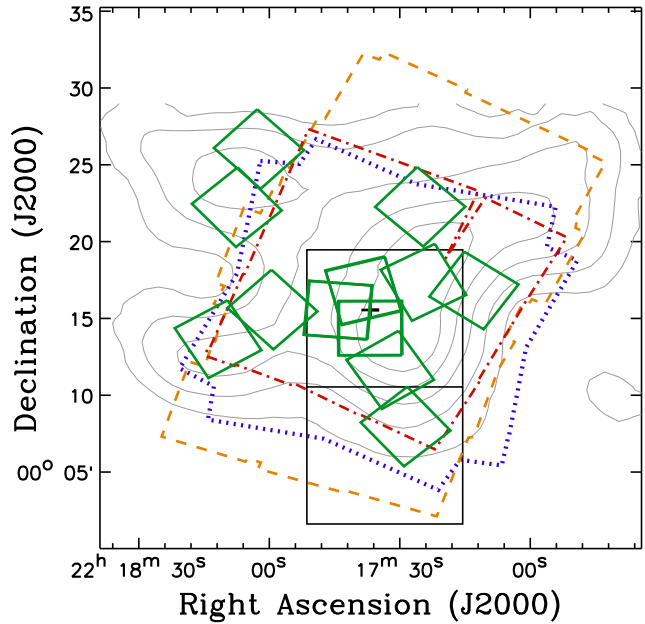}
}
\caption{
Coverage map of SSA22 showing \chandra\ ({\it dotted blue polygon\/}),
\hst\ ({\it green solid squares}), and \spitzer\ IRAC ({\it orange short-dashed
polygon\/}) and MIPS ({\it red dot-dashed polygon\/}) observational regions.
Additionally, we have highlighted the Steidel \etal\ (2003) LBG survey regions
SSA22a and SSA22b ({\it solid black rectangles\/}) and LAE source-density
contours are shown in gray (computed using LAEs from Hayashino \etal\ 2004 and
a spatially-varying density extraction circle with radius of 3.0~arcmin) and have
levels of 0.8, 1.1, 1.4, 1.7, and 2.0 LAEs arcmin$^{-2}$.  
}
\end{figure}

Thus far, we have utilised the \chandra\ and multiwavelength data presented
here to conduct two scientific investigations including (1) Lehmer \etal\
(2009), which finds that the growth of galaxies and SMBHs in the $z = 3.09$
protocluster environment is enhanced by a factor of \hbox{$\approx$2--16} over
that found in $z \approx 3$ field galaxies (i.e., in the CDFs), and (2) Geach
\etal\ (2009), which uses the \chandra\ data to study the host galaxies of LABs
and concludes that AGN and star-formation activity provide more than enough UV
emission to power the extended Ly$\alpha$ emission.  Additional follow-up
investigations and scientific studies are planned and will be presented in
subsequent papers.  

In this paper, we present \chandra\ point-source catalogs and data products
derived from the $\approx$400~ks \hbox{SSA22} data set.  The paper is
orgainized as follows: In $\S$2, we provide details of the observations and
data reduction.  In $\S$3, we present \chandra\ point-source catalogs and
details of our technical analyses.  In $\S$4, we present basic multiwavlength
properties of the \xray\ sources and conduct scientific analyses of sources
within the SSA22 protocluster.  Finally, in $\S$5, we analyse the \xray\
properties of the only obvious extended \xray\ source in the field
\hbox{J221744.6+001738}.  The observational procedures and data processing were
similar in spirit to those presented in Lehmer \etal\ (2005) and Luo \etal\
(2008); however, we have made wider use of the \chandra\ data analysis software
package {\ttfamily ACIS EXTRACT} version~3.131 (Broos \etal\ 2002) while
producing our point-source catalogs (see $\S$3 for further details).  

Throughout this paper, we assume the Galactic column density along the line of
sight to \hbox{SSA22} to be \hbox{$N_{\rm H}=4.6\times 10^{20}$~cm$^{-2}$}
(e.g., Stark et~al.  1992).  The coordinates throughout this paper are J2000.
$H_0$ = 70~\hbox{km s$^{-1}$ Mpc$^{-1}$}, $\Omega_{\rm M}$ = 0.3, and
$\Omega_{\Lambda}$ = 0.7 are adopted throughout this paper (e.g., Spergel
\etal\ 2003), which give the age of the Universe as 13.5~Gyr and imply a
$z=3.09$ look-back time and spatial scale of 11.4~Gyr and
7.6~kpc~arcsec$^{-1}$, respectively.

%
%
%
\begin{table*}
\begin{minipage}{175mm}
\begin{center}
\caption{Journal of SSA22 \chandra\ Observations}
\begin{tabular}{lccccc}
\hline\hline
Obs.                               &
Obs.                                 &
Exposure                             &
\multicolumn{2}{c}{Aim Point}        &
Roll                                 \\
ID                                 &
Start (UT)                         &
Time (ks)$^{\rm a}$                &
$\alpha_{\rm J2000}$                &
$\delta_{\rm J2000}$                &
Angle ($^\circ$)$^{\rm b}$        \\
\hline
9717\dotfill \ldots \ldots  \ldots \ldots \ldots \ldots& 2007-10-01, 06:48 & 116.0 & 22 17 36.80 & +00 15 33.09 & 280.2 \\
8035\dotfill \ldots \ldots  \ldots \ldots \ldots \ldots& 2007-10-04, 04:28 &  96.0 & 22 17 36.80 & +00 15 33.10 & 280.2 \\
8034\dotfill \ldots \ldots \ldots \ldots \ldots \ldots & 2007-10-08, 17:36 & 108.9 & 22 17 36.80 & +00 15 33.10 & 280.2 \\
8036\dotfill \ldots \ldots  \ldots \ldots \ldots \ldots& 2007-12-30, 01:21 &  71.1 & 22 17 37.27 & +00 15 38.37 & 297.9 \\
\hline
\end{tabular}
\end{center}
$^{a}$All observations were continuous.  The short time intervals with bad
satellite aspect are negligible and have not been removed.
$^{b}$Roll angle describes the orientation of the \chandra\ instruments on the
sky.  The angle is between 0--360$^{\circ}$, and it increases to
the West of North (opposite to the sense of traditional position angle).
The total exposure time for the SSA22 observations is 392~ks and the
exposure-weighted average aim point is $\alpha_{\rm J2000} =$~22:17:36.8 and $\delta_{\rm J2000} =$~+00:15:33.1.
\end{minipage}
\end{table*}

%
\section{Observations and Data Reduction}
%

\subsection{Instrumentation and Observations}

The \hbox{ACIS-I} camera (Garmire \etal\ 2003) was used for all of the SSA22
\chandra\ observations.\footnote{For additional information on ACIS and
\chandra\ see the \chandra\ Proposers' Observatory Guide at
http://cxc.harvard.edu/proposer/.} The ACIS-I full field of view is $16.9\times
16.9$~arcmin$^2$ (i.e., \hbox{$31.7 \times 31.7$} comoving Mpc$^2$ at
$z=3.09$), and the sky-projected ACIS pixel size is $\approx$$0.492$~arcsec
($\approx$3.8~kpc pixel$^{-1}$ at $z = 3.09$).  The point-spread function (PSF) is
smallest at the lowest photon energies and for sources with small off-axis
angles and increases in size at higher photon energies and larger off-axis
angles.  For example, the 95~per~cent encircled-energy fraction radius at
1.5~keV at \hbox{0--8~arcmin} off axis is \hbox{$\approx 1.8$--$7.5$~arcsec}
(Feigelson \etal\ 2000; Jerius \etal\ 2000).\footnote{Feigelson et~al.~(2000)
is available on the WWW at
http://www.astro.psu.edu/xray/acis/memos/memoindex.html.} The shape of the PSF
is approximately circular at small off-axis angles, broadens and elongates at
intermediate off-axis angles, and becomes complex at large off-axis angles.

The entire $\approx$400~ks {\it Chandra} exposure consisted of four separate
\chandra\ observations of \hbox{$\approx$70--120~ks} taken between 2007 October
1 and 2007 December 30 and is summarised in Table~1.  The four \hbox{ACIS-I}
CCDs were operated in all of the observations; due to their large angular
offsets from the aim point, the \hbox{ACIS-S} CCDs were turned off for all
observations.  All observations were taken in Very Faint mode to improve the
screening of background events and thus increase the sensitivity of ACIS in
detecting faint \hbox{X-ray} sources.\footnote{For more information on the Very
Faint mode see http://cxc.harvard.edu/cal/Acis/Cal\_prods/vfbkgrnd/ and
Vikhlinin (2001).} Due to dithering and small variations in roll angles and aim
points (see Table~1), the observations cover a total solid angle of 330.0
arcmin$^2$, somewhat larger than a single ACIS-I exposure (295.7 arcmin$^2$).
Combining the four observations gave a total exposure time of 392~ks and an
exposure-weighted average aim point of $\alpha_{\rm J2000} =$~22:17:36.8 and
$\delta_{\rm J2000} =$~+00:15:33.1.

As discussed in $\S$1, the SSA22 \chandra\ target centre was chosen to coincide
with the highest density regions of the $z=3.09$ SSA22 protocluster (see
Fig.~2), where an additional $\approx$79~ks of \hbox{ACIS-S} imaging is already
available (PI: G.~Garmire).  We experimented with source searching utilizing
both the new $\approx$400~ks \hbox{ACIS-I} and previously available
$\approx$79~ks \hbox{ACIS-S} observations and found that the number of detected
sources was not significantly increased.  To avoid the unnecessary
complications of combining data with significantly different aim-points,
backgrounds, and PSFs, we therefore chose to restrict our \chandra\ analyses
to the $\approx$400~ks ACIS-I observations alone.  

\subsection{Data Reduction}

\chandra\ \hbox{X-ray} Center (hereafter CXC) pipeline software was used for
basic data processing, and the pipeline version 7.6.11 was used in all
observations.  The reduction and analysis of the data used \chandra\
Interactive Analysis of Observations ({\ttfamily CIAO}) Version~3.4 tools
whenever possible;\footnote{See http://cxc.harvard.edu/ciao/ for details on
{\ttfamily CIAO}.} however, custom software, including the {\ttfamily TARA}
package, was also used extensively. 

Each level 1 events file was reprocessed using the {\ttfamily CIAO} tool
{\ttfamily acis\_process\_events} to correct for the radiation damage sustained
by the CCDs during the first few months of \chandra\ operations using the
Charge Transfer Inefficiency (CTI) correction procedure of Townsley et~al.
(2000, 2002) and remove the standard pixel randomization.  Undesirable grades
were filtered using the standard \asca\ grade set (\asca\ grades 0, 2, 3, 4, 6)
and known bad columns and bad pixels were removed using a customized
stripped-down version (see $\S$2.2 of Luo \etal\ 2008 for additional details)
of the standard bad pixel file.  The customized bad-pixel file {\it does not}
flag pixels that are thought to have a few extra events per Ms in the
\hbox{0.5--0.7~keV} bandpass.  These events {\it are} flagged as bad pixels and
removed in the standard CXC pipeline-reduced events
lists;\footnote{http://cxc.harvard.edu/cal/Acis/Cal\_prods/badpix/index.html}
however, events landing on these pixels with energies above $\approx$0.7~keV
are expected to be valid events.  We therefore, chose to include these columns
in our analyses. 

After removing the bad pixels and columns, we cleaned the exposures of
cosmic-ray afterglows and hot columns using the {\ttfamily
acis\_detect\_afterglow} procedure.  We found that the use of this procedure
removed cosmic-ray afterglows more stringently than the {\ttfamily
acis\_run\_hotpix} procedure used in the standard CXC reductions without
removing any obvious real \xray\ sources.  

Background light curves for all four observations were inspected using
{\ttfamily EVENT BROWSER} in the Tools for ACIS Real-time Analysis ({\ttfamily
TARA}; Broos et~al. 2000) software package.\footnote{{\ttfamily TARA} is
available at http://www.astro.psu.edu/xray/docs.}  None of the observations had
significant flaring events, defined by the background level being $\simgt$1.5
times higher than nominal.  We therefore did not filter any of the observations
for flaring events.

We registered our observations to the astrometric frame of the UKIRT Infrared
Deep Sky Survey (UKIDSS; e.g., Lawrence \etal\ 2007) Deep Extragalactic Survey
(DXS).  The DXS $K$-band imaging covers the entire \chandra\ observed region of
SSA22 (with total area extending to $\approx$7~deg$^2$) and reaches a 5$\sigma$
limiting magnitude of $K \approx 20.9$ (Vega).  The absolute astrometry for the
DXS imaging has been calibrated using large numbers of Galactic stars
(\hbox{$\approx$60--1000} per observational frame) from the 2MASS database and
source positions are determined to be accurate to $\simlt$0.1~arcsec (see $\S$4
of Lawrence \etal\ 2007).  We ran {\ttfamily wavdetect} at a false-positive
probability threshold of $10^{-5}$ on the events files to create source lists
for each of our four observations (see Table~1).  We then refined the positions
of our four {\ttfamily wavdetect} source lists using PSF centroiding and
matched-filter techniques provided by {\ttfamily ACIS EXTRACT} (see $\S$3.2.1
below).  Using these source lists in combination with the DXS $K$-band catalog,
each aspect solution and events list was registered to the $K$-band frame using
the {\ttfamily CIAO} tools {\ttfamily reproject\_aspect} and {\ttfamily
reproject\_events}, respectively.  The resulting astrometric reprojections gave
nearly negligible linear translations ($<$0.5~pixels), rotations ($<$0.02~deg),
and stretches ($<$0.01~per~cent of the pixel size) for all four observations.
The observation with the smallest reprojections was observation 8034, and we
therefore reprojected the observational frames of observations 8035, 8036, and
9717 to align with observation 8034.  Using the astrometrically-reprojected
events lists, we combined the four observations using {\ttfamily dmmerge} to
create a merged events list.  

\section{Production of Point-Source Catalogs}

\subsection{Image and Exposure Map Creation}

Using the merged events lists discussed in $\S$2.2, we constructed images of
the SSA22 field for three standard bands: \hbox{0.5--8~keV} (full band; FB),
\hbox{0.5--2~keV} (soft band; SB), and \hbox{2--8~keV} (hard band; HB).  These
images have 0.492~arcsec pixel$^{-1}$.  In Figures~3$a$ and 3$b$, we display
the full-band raw and the exposure-corrected adaptively smoothed images (see,
e.g., Baganoff \etal\ 2003), respectively.\footnote{Raw and adaptively smoothed
images for all three standard bands are available at the \hbox{SSA22} website
(see http://astro.dur.ac.uk/$\sim$dma/SSA22/).}  Note that our source detection
analyses have been restricted to the raw images.

We constructed exposure maps for the three standard bands.  These were created
following the basic procedure outlined in $\S$3.2 of Hornschemeier \etal\
(2001) and are normalized to the effective exposures of sources located at the
aim point.  This procedure takes into account the effects of vignetting, gaps
between the CCDs, bad column filtering, and bad pixel filtering.  Also, with
the release of {\ttfamily CIAO} version 3.4, the spatially dependent
degradation in quantum efficiency due to contamination on the ACIS optical
blocking filters is now incorporated into the generation of exposure
maps.\footnote {See http://cxc.harvard.edu/ciao/why/acisqedeg.html}  A photon
index of $\Gamma=1.4$, the slope of the \hbox{X-ray} background in the
\hbox{0.5--8~keV} band (e.g.,\ Marshall \etal\ 1980; Gendreau \etal\ 1995;
Hickox \& Markevitch~2006), was assumed in creating the exposure maps.  The
resulting full-band exposure map is shown in Figure~3$c$.  Figure~4$a$ displays
the SSA22 survey solid angle as a function of effective exposure for the three
standard bands.  The majority of the $\approx$330~arcmin$^2$ field
($\approx$53--69~per~cent depending on the bandpass) has an effective exposure
exceeding 300~ks.

%
%

\begin{figure*}
\centerline{
\includegraphics[width=16cm]{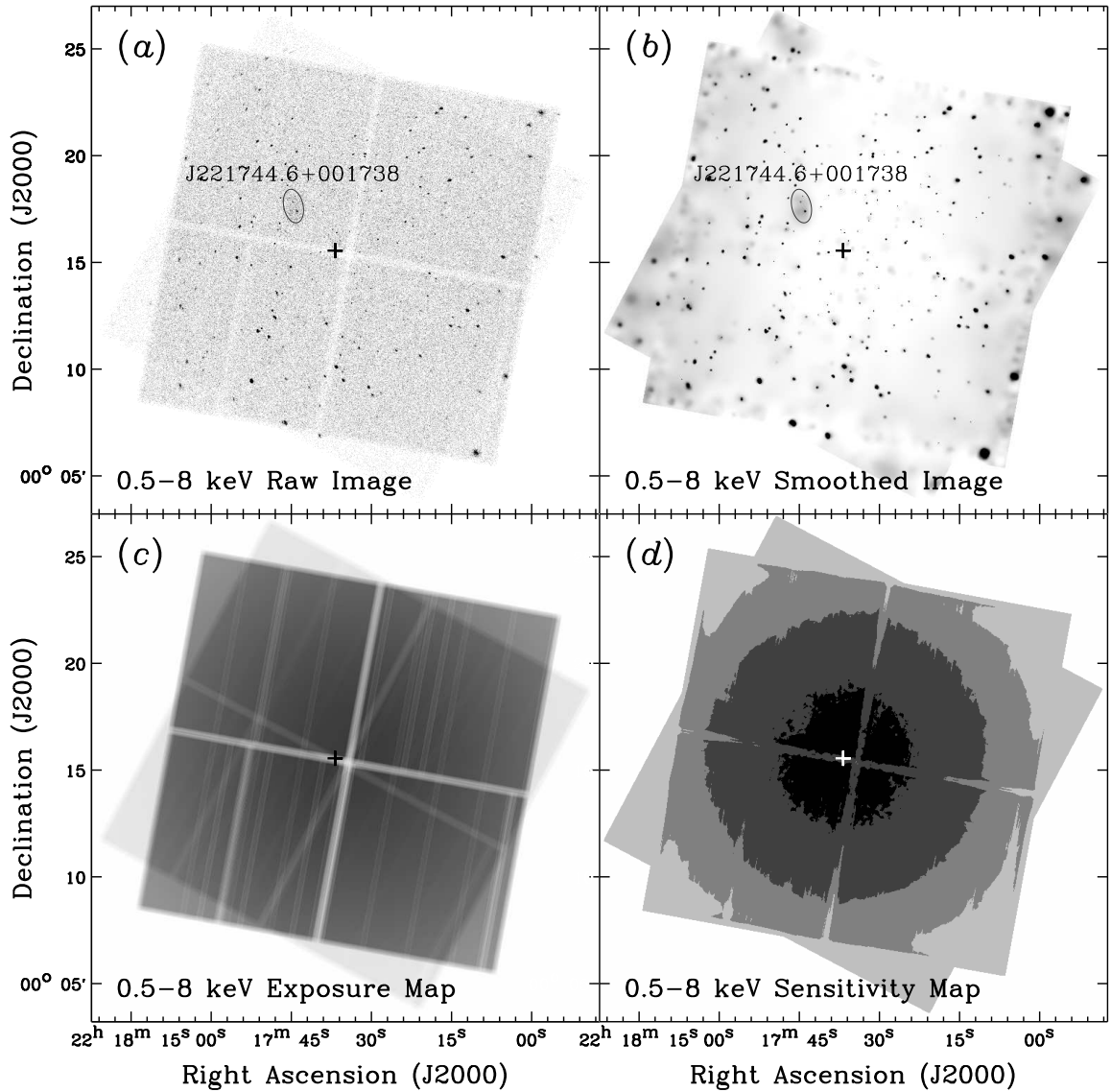}
}
\caption{
({\it a\/}) Full-band (0.5--8~keV) raw image of the $\approx$400~ks
\hbox{SSA22} field.  This image was constructed following the procedure
outlined in $\S$3.1 and has been binned by a factor of four in both right
ascension and declination.  The location of the exposure-weighted average aim
point has been shown with a cross symbol and the extended source
\hbox{J221744.6+001738} ellipse has been outlined (see $\S$5 for discussion).  
({\it b\/}) Adaptively-smoothed exposure-corrected full-band image of
\hbox{SSA22}.  This image was created following the techniques discussed in
$\S$3.1.  
({\it c\/}) Full-band exposure map of SSA22.  The exposure map was created
following the procedure outlined in $\S$3.1.  The grayscales are linear with
the darkest areas corresponding to the highest effective exposure times (with a
maximum pixel value of $\approx$388~ks).  Note the chip gaps in running between
the four ACIS-I CCDs. 
({\it d\/}) Full-band sensitivity map of SSA22.  This sensitivity map has been
created following the procedure outlined in $\S$3.3.  The gray-scale levels
(running from black to light gray) indicate regions with flux limits (in units
of \flux) of $< 2 \times 10^{-16}$, 2--5~$\times$~$10^{-16}$,
5--10~$\times$~$10^{-16}$, and $> 10^{-15}$, respectively.  
}
\end{figure*}

\subsection{Point-Source Searching and Catalog Production}

Our point-source catalog production has been tailored to generate source lists
that can be directly comparable with those from previous studies of the CDFs
(e.g., Alexander \etal\ 2003; Lehmer \etal\ 2005; Luo \etal\ 2008) to enable
comparative studies; however, our procedure and main catalog definitions differ
in a number of important ways.  The main differences in the catalog production
procedure adopted here are as follows:

\vspace{0.1in}

1.---We first created a {\it candidate-list catalog} of sources detected by
{\ttfamily wavdetect} at a liberal false-positive probability threshold of
10$^{-5}$.  We then created a more conservative {\it main catalog}, in which we
evaluated the significance of each detected source candidate (see point 2
below) and included only \xray\ sources that had high statistical probabilities
of being true sources considering their local backgrounds.  This approach
produced point-source catalogs that are of similar quality to those produced by
running {\ttfamily wavdetect} at the more typical false-positive probability
threshold of $10^{-6}$, but allowed for flexibility in the inclusion of
additional legitimate sources that fell below the $10^{-6}$ threshold (see
$\S$3.2.1 for further details).  

2.---While computing \xray\ source properties and evaluating the significance
of each detected source (i.e., the probability of it being a true source), we
made wide use of the {\ttfamily ACIS EXTRACT} (hereafter, {\ttfamily AE})
point-source analysis software.\footnote{The {\ttfamily ACIS EXTRACT} software
is available on the WWW at
http://www.astro.psu.edu/xray/docs/TARA/ae\_users\_guide.html} \ae\ contains
algorithms that allow for appropriate computation of source properties when
multiple observations with different roll angles and/or aim points are being
combined and analysed (see discussion below for further details).  Our adopted
procedure has previously been utilised in similar forms by other groups,
including, e.g., the \xray\ source catalogs produced in the \chandra\ Orion
ultra-deep project (COUP; e.g., Getman \etal\ 2005) and the \aegisx\ survey
(e.g., Nandra \etal\ 2005; Laird \etal\ 2009).

\subsubsection{Candidate-List Catalog Production}

We began by generating a {\it candidate-list catalog} of \xray\ point sources
using the {\ttfamily CIAO} tool {\ttfamily wavdetect} (Freeman \etal\ 2002).
We performed our searching in the three standard band (i.e., the FB, SB, and
HB) images using a ``$\sqrt{2}$~sequence'' of wavelet scales (i.e.,\ 1,
$\sqrt{2}$, 2, $2\sqrt{2}$, 4, $4\sqrt{2}$, and 8 pixels) and a false-positive
probability threshold of $10^{-5}$.  We note that the use of a false-positive
probability threshold of $10^{-5}$ is expected to generate a non-negligible
number of spurious sources with low source counts \hbox{($\simlt$2--3)};
however, as noted by Alexander \etal\ (2001), real sources can be missed using
a more stringent source-detection threshold (e.g., $10^{-6}$).  In $\S$3.2.2
below, we create a main catalog of sources by determining the significance of
each source in our candidate-list catalog and excising sources with individual
detection significances falling below an adopted threshold.

%
%
\begin{figure*}
\centerline{
\includegraphics[width=9cm]{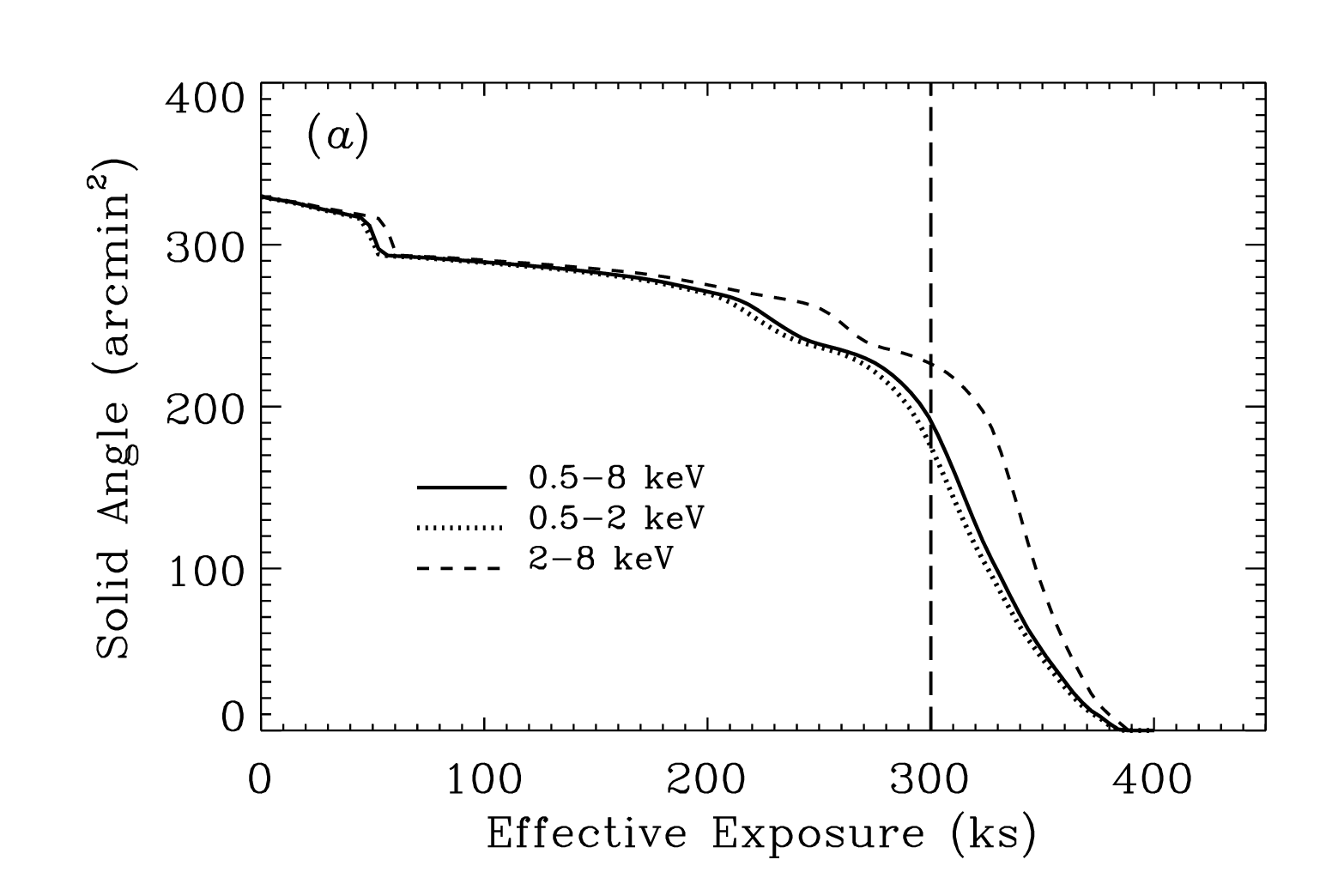}
\hfill
\includegraphics[width=9cm]{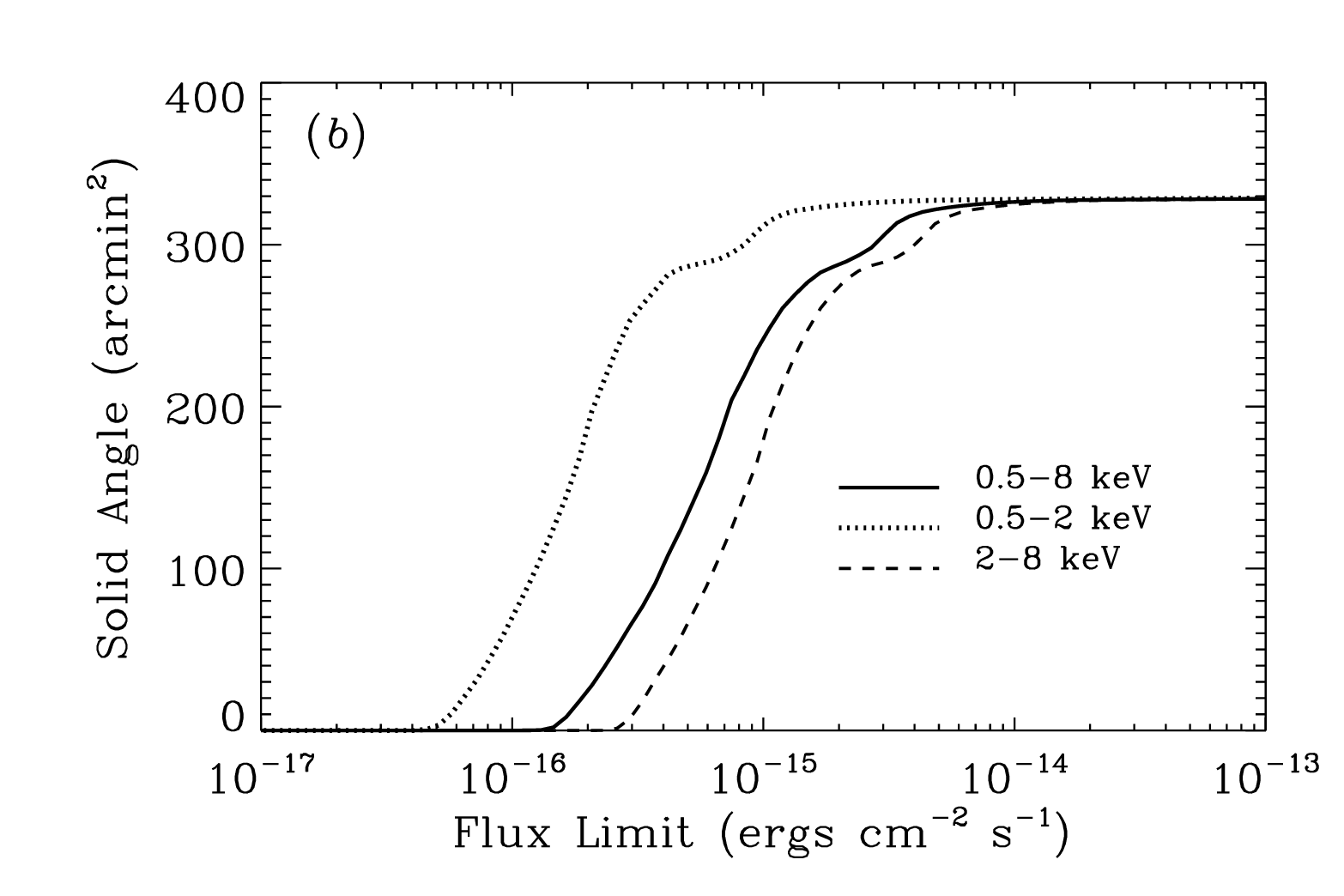}
}
\caption{
({\it a\/}) Total solid angle with at least the vignetting-corrected effective
exposure indicated on the abscissa for each of the three standard bands.  We
find that $\simgt$60~per~cent of the entire $\approx$330~arcmin$^2$ field has an
effective exposure exceeding 300~ks ({\it vertical long-dashed line\/}).
({\it b\/}) Solid angle versus flux limit for the three standard bands.  These
curves were computed using the sensitivity maps described in $\S$3.3.  In the
most sensitive region of the SSA22 field, within the most sensitive
10~arcmin$^2$ region (predominantly around the \chandra\ aim point), the flux
limits are $\approx$$1.7 \times 10^{-16}$ (full band), $\approx$$5.7 \times
10^{-17}$ (soft band), and $\approx$$3.0 \times 10^{-16}$ (hard band) \flux.
}
\end{figure*}

Our candidate-list catalog (constructed using a false-positive probability
threshold of $10^{-5}$) contained a total of 350 \xray\ source candidates.  For
this candidate list, we required that a point source be detected in at least
one of the three standard bands with {\ttfamily wavdetect}.  We utilise
full-band source positions for all sources with full band detections; for
sources not detected in the full band, we utilised, in order of priority, soft
band and hard band source positions.  Cross-band matching was performed using a
2.5~arcsec matching radius for sources within 6.0~arcmin of the
exposure-weighted mean aim point and 4.0~arcsec for sources at off-axis angles
$>$6.0~arcmin.  These matching radii were chosen based on inspection of
histograms showing the number of matches obtained as a function of angular
separation (e.g., see \S2 of Boller \etal\ 1998); with these radii, the
mismatch probability is expected to be $\simlt$1~per~cent over the entire
field.  While matching the three standard band source lists, we found that no
source in one band matched to more than one source in another band.

Using \ae, we improved the {\ttfamily wavdetect} source positions using
centroiding and matched-filter techniques.  The matched-filter technique
convolves the full-band image in the vicinity of each source with a combined
PSF.  The combined PSF is produced by combining the ``library'' PSF of a source
for each relevant observation (from Table~1), weighted by the number of
detected counts.\footnote{The PSFs are taken from the CXC PSF library, which is
available on the WWW at http://cxc.harvard.edu/ciao/dictionary/psflib.html.}
This technique takes into account the fact that, due to the complex PSF at
large off-axis angles, the \xray\ source position is not always located at the
peak of the \xray\ emission.  For sources further than 8~arcmin from the
average aim-point, the matched-filter technique provides positions that have
median offsets to potential $K$-band counterparts (within a matching radius of
1.5~arcsec) that are $\approx$0.1~arcsec smaller than such offsets found for both
centroid and {\ttfamily wavdetect} source positions.  For sources with off-axis
angles $\theta$~$<$~8~arcmin, we found that the centroid positions had the
smallest median offsets to $K$-band counterparts compared with matched-filter
and {\ttfamily wavdetect} positions.  Thus, for our candidate-list source
positions, we utilised centroid positions for $\theta$~$<$~8~arcmin and
matched-filter positions for $\theta$~$\ge$~8~arcmin.  The median source
position shift relative to {\ttfamily wavdetect} source positions was
$\approx$0.26~arcsec (quartile range of $\approx$0.1--0.5~arcsec).  

For our candidate-list catalog sources, we performed photometry using \ae.  For
each \xray\ source in each of the four observations listed in Table~1, \ae\
extracted full band source events and exposure times (using the events lists
and exposure maps discussed in $\S$2.2 and $\S$3.1, respectively) from all
pixels that had exposure within polygonal regions centred on the \xray\
position.  Each polygonal region was constructed by tracing the
$\approx$90~per~cent encircled-energy fraction contours of a local PSF measured
at 1.497~keV that was generated using {\ttfamily CIAO} tool {\ttfamily mkpsf}.
Upon inspection, we found 16 sources (i.e., eight source pairs) had
significantly overlapping polygonal extraction regions.  For these sources, we
utilised smaller extraction regions (ranging from \hbox{30--75~per~cent}
encircled energy fractions), which were chosen to be as large as possible
without overlapping.  For each source, the source events and exposure times
from all relevant observations were then summed to give the total source events
$S_{\rm src}$ and exposure times $T_{\rm src}$.

For each source in our candidate-list catalog, local full band background
events and exposure times were then extracted from each observation.  This was
achieved by creating full band events lists and exposure maps with all
candidate-list catalog sources masked out of circular masking regions of radius
$1.1 \times$~the size of the $\approx$99.9~per~cent encircled energy fraction
at 1.497~keV (estimated using each local PSF).  The radii of the masked out
regions cover a range of \hbox{$\approx$11.6--17.9~arcsec} and have a median
value of $\approx$12.8~arcsec.  Using these masked data products, \ae\ then
extracted events and exposure times from a larger background-extraction
circular aperture that was centred on the source.  The size of the background
extraction radius varied with each source and was chosen to encircle
$\approx$170 full-band (resulting in $\approx$40 soft-band and $\approx$130
hard-band) background events for each observation.  The background extraction
radius is therefore affected by the masking of nearby sources and could
potentially become large and non-local for the case of crowded regions.
However, our candidate-list catalog has negligible source crowding and
therefore the radii of our background extraction regions remain relatively
small (\hbox{$\approx$22--55 arcsec}) and are therefore representative of the
local backgrounds.  For each source, the background events and exposure times
from all relevant observations were then summed to give total background events
$S_{\rm bkg}$ and exposure times $T_{\rm bkg}$.

The source and local background events extracted above were then filtered by
photon energy to produce source and local background counts appropriate for
each of the three standard bands.  For each bandpass, net source counts
\hbox{$N^{E_1-E_2}$} (where $E_1$ and $E_2$ represent the lower and upper
energy bounds, respectively, of an arbitrary bandpass) were computed following
\hbox{$N^{E_1-E_2} = [S_{\rm src}^{E_1-E_2} - S_{\rm bkg}^{E_1-E_2} T_{\rm
src}/T_{\rm bkg}]/\gamma_{\rm EEF}^{E_1-E_2}$,} where $\gamma_{\rm
EEF}^{E_1-E_2}$ is the encircled-energy fraction appropriate for the source
extraction region and bandpass.  For the soft and hard bands, $\gamma_{\rm
EEF}^{\rm 0.5-2~keV}$ and $\gamma_{\rm EEF}^{\rm 2-8~keV}$ were approximated
using PSF measurements at 1.497~keV and 4.51~keV, respectively.  For the full
band, we used the approximation \hbox{$\gamma_{\rm EEF}^{\rm 0.5-8~keV} \approx 1/2
(\gamma_{\rm EEF}^{\rm 0.5-2~keV} + \gamma_{\rm EEF}^{\rm 2-8~keV} )$}.  As a
result of our liberal searching criteria used for generating the candidate-list
catalog, many sources have small numbers of counts.  We find that 51 (3)
candidate-list catalog sources have $<$5 ($<$0) net counts in all three of the
standard bandpasses.  In the next section, we evaluate the reliability of the
sources detected in our candidate-list catalog on a source-by-source basis to
create a more conservative {\it main catalog} of reliable sources.

\subsubsection{Main Chandra Source Catalog Selection}

As discussed above, our candidate-list catalog of 350 \xray\ point-sources
produced by running {\ttfamily wavdetect} at a false-positive probability
threshold of $10^{-5}$ is expected to have a significant number of false
sources.  If we conservatively treat our three standard band images as being
independent, it appears that $\approx$130 ($\approx$37~per~cent) false sources are
expected in our candidate-list \chandra\ source catalog for the case of a
uniform backgrounds over $\approx$$1.3 \times 10^7$ pixels (i.e., all three
bands).  However, since {\ttfamily wavdetect} suppresses fluctuations on scales
smaller than the PSF, a single pixel usually should not be considered a source
detection cell, particularly at large off-axis angles.  Hence, our false-source
estimates are conservative.  As quantified in $\S$3.4.1 of Alexander \etal\
(2003) and by source-detection simulations (P.~E.~Freeman 2005, private
communication), the number of false-sources is likely \hbox{$\approx$2--3}
times less than our conservative estimate, leaving only \hbox{$\approx$40--65}
false sources; still potentially a significant fraction (i.e., $\simlt$19~per~cent) of
our candidate-list list of 350 point sources.  

In order to produce a more conservative {\it main catalog} of \chandra\
point-sources and remove likely false sources, we evaluated for each source the
binomial probability $P$ that no source exists given the measurements of the
source and local background (see $\S$3.2.1 for details on the measurements of
source and local background events).  The quantity $P$ is computed using \ae\
for each of the three standard bands (see also Appendix~A2 of Weisskopf \etal\
2007 for further details).  For a source to be included in our main catalog, we
required $P < 0.01$ in at least one of the three standard bands.  This
criterion gave a total of 297 sources, which make up our main catalog (see
$\S$3.2.3 and Table~2 below).

%
%
\begin{figure}
\centerline{
\includegraphics[width=9cm]{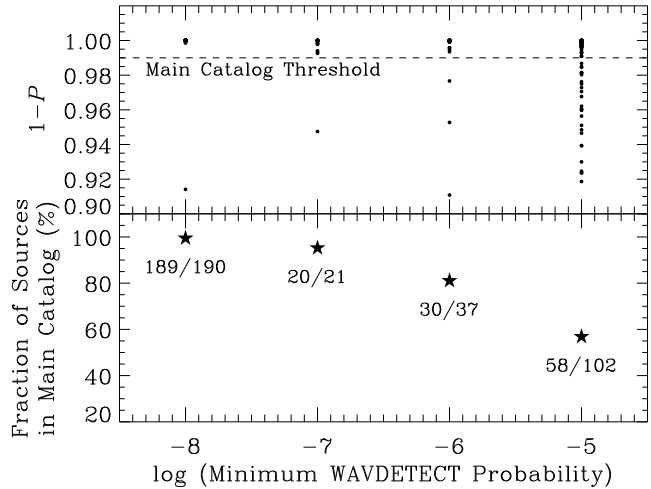}
}
\caption{
({\it top\/}) One minus the \ae\ binomial probability that no source exists
($1-P$) versus the logarithm of the minimum {\ttfamily wavdetect} probability
for the 350 sources in our candidate-list catalog ({\it small filled circles\/}; see
$\S\S$~3.2.1 and 3.2.2).  We note that for a given value of the {\ttfamily
wavdetect} probability, the majority of the sources have $1-P = 1$, causing
clumping of the symbols.  The dashed horizontal line at $1-P = 0.99$ shows the 
adopted lower threshold used for including a source in our main catalog.
({\it bottom\/}) The percentage of sources in the candidate-list catalog with $1-P >
0.99$, which were included in our main catalog, as a function of minimum
{\ttfamily wavdetect} probability ({\it five-pointed stars\/}).  The number of
sources with $1-P > 0.99$ versus the number of candidate-list catalog sources detected
at each minimum {\ttfamily wavdetect} probability has been annotated in the
figure.  We find that the fraction of candidate-list catalog sources included in our
main catalog falls from $\approx$99.5~per~cent to $\approx$56.9~per~cent between minimum
wavedetect probabilities of $10^{-8}$ and $10^{-5}$.
}
\end{figure}

Our adopted binomial probability-based detection criterion has a number of
advantages over a direct {\ttfamily wavdetect} approach including (1) the more
detailed treatment of complex source extraction regions for exposures with
multiple observations that have different aim points and roll angles, (2)
better source position determination before count extraction and probability
measurements are made, and (3) a more transparent mathematical criterion (i.e.,
the binomial probability) that is used for the detection of a source.  However,
this approach has the disadvantage that we do not approximate the shape of the
PSF, as is done by {\ttfamily wavdetect} through the use of the ``Mexican Hat''
source detection wavelet.  As we will highlight below, our adopted procedure
recovers almost all of the sources detected with {\ttfamily wavdetect} at a
false-positive probability threshold of $10^{-6}$ and a large number of
additional real sources detected at $10^{-5}$; therefore, this procedure is
preferred over a direct {\ttfamily wavdetect} approach.

To give a more detailed {\ttfamily wavdetect} perspective on source
significance, we ran {\ttfamily wavdetect} over the three standard band images
at additional false-positive probability thresholds of $10^{-6}$, $10^{-7}$,
and $10^{-8}$ and found detections for 248 ($\approx$71~per~cent), 211
($\approx$60~per~cent), and 190 ($\approx$54~per~cent) of the 350
candidate-list catalog sources, respectively.  Out of the 53 sources that were
excluded from our candidate-list catalog to form our main catalog (297 sources;
down from the 350 sources in our candidate-list catalog), 9 had
{\ttfamily wavdetect} false-positive probability detection thresholds $\le$$10^{-6}$.
For convenience, in Appendix~A we present the properties of the nine sources
with {\ttfamily wavdetect} detection threshold $\le$$10^{-6}$ that were not
included in the main catalog.  This provides the option for the reader to
construct a pure {\ttfamily wavdetect} catalog down to a false-positive
probability threshold of $10^{-6}$, $10^{-7}$, or $10^{-8}$, to make it more
consistent with previous \chandra\ source catalogs.

In Figure~5, we present the \ae-determined binomial probabilities and the
fraction of sources included in the main catalog as a function of the minimum
{\ttfamily wavdetect} probability for the 350 candidate-list catalog sources.
Our main catalog includes 58 sources that had minimum {\ttfamily wavdetect}
probabilities of $10^{-5}$.  For these 58 sources, we performed cross-band
matching between the \xray\ source positions and the $K$-band and \spitzer\
3.6, 4.5, 5.8, and 8.0$\mu$m IRAC source positions.  In total, we find that 33
($\approx$56.9~per~cent) of the 58 sources had at least one infrared
counterpart within a 1.5~arcsec matching radius; by comparison, we find that
the infrared counterpart fraction for the remaining main \chandra\ catalog
sources (i.e., those with minimum {\ttfamily wavdetect} probabilities
$\le$$10^{-6}$) is $\approx$89.1~per~cent.  We estimated the expected number of
false matches by shifting the main \chandra\ catalog source positions by a
constant offset (5.0~arcsec) and rematching them to the infrared positions.  We
performed four such ``shift and rematch'' trials in unique directions and found
that on an average $\approx$8.2$^{+2.1}_{-1.7}$~per~cent of the shifted source
positions had an infrared match.  We therefore estimate that of the 58 main
catalog sources that had minimum {\ttfamily wavdetect} probabilities of
$10^{-5}$, at least \hbox{$\approx$27--29}
(\hbox{$\approx$46.6--50.4~per~cent}) have true infrared counterparts that are
associated with the \xray\ sources; we note that since these are typically
fainter X-ray sources we might expect a lower matching fraction than that found
for the brighter X-ray sources.  This analysis illustrates that our main
catalog selection criteria effectively identifies a significant number of
additional {\it real} \xray\ sources below the traditional $10^{-6}$ {\ttfamily
wavdetect} searching threshold through \xray\ selection alone.

\subsubsection{Properties of Main Catalog Sources}

To estimate positional errors for all 297 sources in our main \chandra\
catalog, we performed cross-band matching between the \xray\ and $K$ bands.  As
discussed in $\S$2.2, the $K$-band sources have highly accurate and precise
absolute astrometric positions ($\simlt$0.1~arcsec positional errors) and reach a
5$\sigma$ limiting magnitude of \hbox{$K \approx 20.9$} (Vega), a regime where
the source density is relatively low ($\approx$80,000~sources~deg$^{-2}$) and
therefore ideal for isolating highly-confident near-infrared counterparts to
\xray\ sources with little source confusion.  Using a matching radius of
2.5~arcsec, we find that 193 ($\approx$65~per~cent) \hbox{X-ray} sources have $K$-band
counterparts down to $K \approx 20.9$.  For 13 of our main catalog sources,
there was more than one $K$-band counterpart within 2.5~arcsec.  In these cases,
we chose the source with the smallest offset as being the most likely
counterpart.  We computed the expected number of false matches using the shift
and rematch technique described in $\S$3.2.2 and estimate that $\approx$26.3
($\approx$14~per~cent) of the matches are false (with a random offset of 1.72~arcsec).
We note that our choice to use the large 2.5~arcsec matching radius is based on
the fact that the \chandra\ positional errors are expected to be of this order
for low-count sources that are far off-axis where the PSF is large.  A more
conservative matching criterion is used to determine likely $K$-band
counterparts, which are reported in the main catalog (see description of
columns~[17] and [18] of Table~2 below).  In a small number of cases, the
\hbox{X-ray} source may be offset from the centroid of the $K$-band source even
though both are associated with the same galaxy (e.g., a $z \simlt 0.3$ galaxy
with extended $K$-band emission from starlight that also has an off-nuclear
ultraluminous \hbox{X-ray} binary with
\hbox{\Lx~$\approx$~10$^{38}$--$10^{40}$~\xlum;} see e.g., Hornschemeier \etal\
2004; Lehmer \etal\ 2006).  

In Figure~6, we show the positional offset between the \hbox{X-ray} and
$K$-band sources versus off-axis angle.  The median offset is
$\approx$0.36~arcsec; however, there are clear off-axis angle and source-count
dependencies.  The off-axis angle dependence is due to the PSF becoming
broad at large off-axis angles, while the count dependency is due to the fact
that sources having larger numbers of counts provide a better statistical
sampling of the local PSF.  To estimate the positional errors of our sources
and their dependencies, we implemented the parameterization provided by Kim
\etal\ (2007), derived for sources in the \chandra\ Multiwavelength Project
(ChaMP):

\begin{equation}
\log \Delta_{\rm X} = a_0 + a_1 \theta + a_2 \log C,
\end{equation}

\noindent where $\Delta_{\rm X}$ is the positional error in arcseconds,
$\theta$ is the off-axis angle in units of arcminutes, $C$ is the net counts
from the energy band used to determine the position, and $a_0$, $a_1$, and
$a_2$ are constants.  In equation~1, $a_0$ provides a normalization to the
positional error and $a_1$ and $a_2$ indicate the respective off-axis angle and
source-count dependencies.  Initial values of $a_0$, $a_1$, and $a_2$ were
determined by performing multivariate $\chi^2$ minimization of equation~1,
using our main catalog and the \xray/$K$-band offsets as a proxy for the
positional error.  Using the resulting best-fit equation, the value of $a_0$
was subsequently adjusted upward by a constant value until $\simgt$80~per~cent
of the main catalog sources with $K$-band counterparts had $\Delta_{\rm X}$
values that were larger than their \xray/$K$-band offsets.  This resulted in
values of $a_0 = -0.0402$, $a_1 = 0.0960$, and $a_2 = -0.3542$; applying these
values to equation~1 gives positional errors with $\approx$80~per~cent
confidence.  Note that the values of $a_1$ and $a_2$ are similar to those found
from ChaMP (see equation~12 of Kim \etal\ 2007); however, due to differences in
adopted confidence levels ($\approx$95~per~cent for ChaMP) and the positional
refinements described in $\S$3.2.1, our value of $a_0$ is smaller by
$\approx$0.4~dex.

%
%
\begin{figure}
\centerline{
\includegraphics[width=9.5cm]{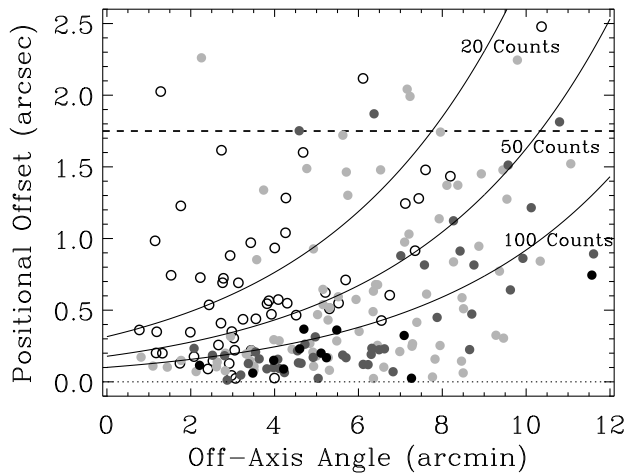}
}
\caption{
Positional offset versus off-axis angle for sources in the main {\it Chandra}
catalog that were matched to near-infrared sources from the UKIDSS DXS $K$-band
image to within 2.5~arcsec.  Symbol grayscales indicate various ranges of net
source counts including $<$20 ({\it open circles\/}), 20--50 ({\it light
gray\/}), 50--100 ({\it dark gray\/}), and $>$100 ({\it filled black
circles\/}) net counts.  Our $\approx$80~per~cent positional error curves, computed
following equation~1, have been indicated for sources with 20, 50, and 100 net
counts.  For reference, we have highlighted the median offset (1.72~arcsec; {\it
horizontal dashed line\/}) found for source matching to random positions; this
was obtained using the shift and rematch technique described in $\S$3.2.2.
}
\end{figure}

The main {\it Chandra} source catalog is presented in Table~2, and the details
of the columns are given below.

\vspace{0.2in}

Column~(1) gives the source number. Sources are listed in order of increasing
right ascension.  Source positions were determined following the procedure
discussed in $\S$3.2.1.

Columns~(2) and (3) give the right ascension and declination of the
\hbox{X-ray} source, respectively.  To avoid truncation error, we quote the
positions to higher precision than in the International Astronomical Union
(IAU) registered names beginning with the acronym ``CXO SSA22'' for ``\chandra\
\hbox{X-ray} Observatory Small Selected Area 22.'' The IAU names should be
truncated after the tenths of seconds in right ascension and after the
arcseconds in declination.

Column~(4) and (5) give the \ae\ significance, presented as unity minus the
computed binomial probability $P$ that no \xray\ source exists ($1-P$), and the
logarithm of the minimum false-positive probability run with {\ttfamily
wavdetect} in which each source was detected, respectively.  Lower values of
the binomial probability and false-positive probability threshold indicate a
more significant source detection.  We find that 189, 20, 30, and 58 sources
have {\it minimum} {\ttfamily wavdetect} false-positive probability thresholds
of $10^{-8}$, $10^{-7}$, $10^{-6}$, and $10^{-5}$, respectively.

Column~(6) gives the $\approx$80~per~cent positional uncertainty in arcseconds,
computed following equation~1 (see above), which is dependent on the off-axis
angle and the net counts of the source in the detection band used to determine
the photometric properties (see columns~[8]--[16]).

Column~(7) gives the off-axis angle for each source in arcminutes.  This is
calculated using the source position given in columns~(2) and (3) and the
exposure-weighted mean aim point.

Columns~(8)--(16) give the net background-subtracted source counts and the
corresponding $1\sigma$ upper and lower statistical errors (from Gehrels 1986),
respectively, for the three standard bands.  Source counts and statistical
errors have been calculated by \ae\ using the position given in columns~(2) and
(3) for all bands and following the methods discussed in detail in $\S$3.2.1,
and have not been corrected for vignetting.  We note that the extraction of
source counts and the computation of statistical errors was performed for all
sources in our candidate-list catalog of sources detected using {\ttfamily
wavdetect} at a false-positive probability threshold of $1 \times 10^{-5}$.
Since all {\it candidate-list catalog} sources were masked when calculating
local backgrounds for our {\it main catalog} sources (see $\S$3.2.1), this
could in principle have a mild effect on background calculations in cases where
lower-significance candidate-list catalog sources (i.e., sources that would
later not be included in the main catalog) were near main catalog sources.  We
note, however, that since the number of candidate-list catalog sources that
were excluded from our main catalog is small (i.e., 53 sources), there is
little overlap between these sources and the background extraction regions of
our main catalog.  We found that 43 ($\approx$14~per~cent) of the 297 main
catalog background extraction regions had some overlap with the
$\approx$90~per~cent PSF regions of at least one of the 53 excluded
candidate-list catalog sources.  Since the 53 excluded candidate-list catalog
sources already have count estimates that were consistent with the local
background, we conclude that this will not have a significant effect on our
main catalog source properties.

To be consistent with our point-source detection criteria defined in
$\S$3.2.2, we considered a source to be ``detected'' for photometry purposes
in a given band if the \ae-computed binomial probability for that band has a
value of $P < 0.01$.  When a source is not detected in a given band, an upper
limit is calculated; these sources are indicated as a ``$-$1'' in the error
columns.  All upper limits were computed using the \ae-extracted photometry
(see $\S$3.2.1) and correspond to the 3$\sigma$ level appropriate for Poisson
statistics (Gehrels~1986).

%
%

\begin{table*}
\begin{minipage}{175mm}
\begin{center}
\caption{Main \chandra\ Catalog}
\begin{tabular}{cccccccccc}
\hline\hline
 & \multicolumn{2}{c}{X-ray Coordinates}  & \multicolumn{2}{c}{Detection Probability} &  & & \multicolumn{3}{c}{Net Counts} \\
Source & \multicolumn{2}{c}{\rule{1.2in}{0.01in}} & \multicolumn{2}{c}{\rule{1.2in}{0.01in}} & Pos. Error & $\theta$ & \multicolumn{3}{c}{\rule{1.8in}{0.01in}}  \\
 Number & $\alpha_{\rm J2000}$ & $\delta_{\rm J2000}$  & {\ttfamily AE} & {\ttfamily wavdetect} & (arcsec) & (arcmin) & 0.5--8~keV & 0.5--2~keV & 2--8~keV  \\
 (1) & (2) & (3) & (4) & (5) & (6) & (7) & (8)--(10) & (11)--(13) & (14)--(16) \\
\hline
   1 &     22 16 51.96 &     +00 18 49.0 &  1.000 & $-$8 &  3.60 & 11.77 &          32.1$^{+10.0}_{-8.8}$ &           16.5$^{+6.4}_{-5.2}$ &           15.6$^{+8.4}_{-7.2}$\\
   2 &     22 16 55.25 &     +00 21 54.2 &  1.000 & $-$8 &  3.37 & 12.22 &         51.1$^{+12.7}_{-11.5}$ &           32.8$^{+8.5}_{-7.3}$ &                        $<$30.3\\
   3 &     22 16 56.32 &     +00 16 57.7 &  1.000 & $-$8 &  2.27 & 10.34 &          48.0$^{+10.6}_{-9.4}$ &           39.1$^{+8.3}_{-7.0}$ &                        $<$22.1\\
   4 &     22 16 58.20 &     +00 21 58.6 &  1.000 & $-$8 &  1.35 & 11.61 &        458.8$^{+28.2}_{-27.0}$ &        249.7$^{+19.2}_{-18.1}$ &        208.5$^{+21.3}_{-20.1}$\\
   5 &     22 16 58.19 &     +00 18 55.1 &  1.000 & $-$8 &  2.21 & 10.22 &         48.3$^{+16.0}_{-14.8}$ &           19.0$^{+8.7}_{-7.5}$ &         29.5$^{+14.1}_{-12.9}$\\
   6 &     22 16 59.08 &     +00 15 13.4 &  1.000 & $-$8 &  1.44 &  9.57 &        107.1$^{+13.4}_{-12.2}$ &          75.6$^{+10.7}_{-9.5}$ &           30.6$^{+8.9}_{-7.6}$\\
   7 &     22 17 00.33 &     +00 19 55.2 &  1.000 & $-$8 &  1.30 & 10.12 &        201.9$^{+21.0}_{-19.9}$ &        127.0$^{+14.4}_{-13.2}$ &         73.9$^{+16.1}_{-14.8}$\\
   8 &     22 17 00.50 &     +00 21 23.7 &  1.000 & $-$8 &  1.43 & 10.80 &        236.1$^{+21.8}_{-20.6}$ &        158.5$^{+15.7}_{-14.5}$ &         76.2$^{+15.9}_{-14.6}$\\
   9 &     22 17 02.23 &     +00 13 09.5 &  0.994 & $-$7 &  2.08 &  9.00 &         26.8$^{+13.3}_{-12.1}$ &                        $<$19.5 &         26.0$^{+12.2}_{-11.0}$\\
  10 &     22 17 03.00 &     +00 15 25.7 &  0.997 & $-$5 &  1.76 &  8.47 &         30.7$^{+13.0}_{-11.8}$ &                        $<$21.5 &                        $<$34.5\\
\\
  11 &     22 17 04.90 &     +00 09 39.3 &  1.000 & $-$8 &  1.06 &  9.92 &        315.8$^{+23.6}_{-22.4}$ &        145.0$^{+15.0}_{-13.8}$ &        171.3$^{+18.9}_{-17.7}$\\
  12 &     22 17 05.41 &     +00 15 14.0 &  1.000 & $-$8 &  0.67 &  7.88 &        320.7$^{+21.8}_{-20.6}$ &        208.6$^{+16.9}_{-15.7}$ &        110.0$^{+14.4}_{-13.2}$\\
  13 &     22 17 05.63 &     +00 19 46.3 &  1.000 & $-$6 &  1.80 &  8.87 &         37.2$^{+12.4}_{-11.2}$ &                        $<$17.8 &         35.3$^{+11.5}_{-10.3}$\\
  14 &     22 17 05.82 &     +00 22 27.7 &  0.993 & $-$5 &  3.68 & 10.37 &                        $<$46.3 &          12.6$^{+11.1}_{-7.0}$ &                        $<$39.7\\
  15 &     22 17 05.83 &     +00 22 24.7 &  0.998 & $-$7 &  2.74 & 10.34 &         28.2$^{+16.4}_{-12.4}$ &                        $<$32.2 &                        $<$44.7\\
  16 &     22 17 06.14 &     +00 13 38.0 &  1.000 & $-$8 &  1.12 &  7.93 &         79.4$^{+13.6}_{-12.4}$ &           39.0$^{+8.9}_{-7.6}$ &          40.4$^{+11.1}_{-9.9}$\\
  17 &     22 17 06.69 &     +00 18 38.5 &  1.000 & $-$6 &  1.71 &  8.15 &          27.3$^{+10.7}_{-9.5}$ &           19.7$^{+7.2}_{-6.0}$ &                        $<$25.9\\
  18 &     22 17 07.04 &     +00 14 29.8 &  0.997 & $-$5 &  1.63 &  7.53 &           21.4$^{+9.9}_{-8.7}$ &                        $<$17.3 &                        $<$26.4\\
  19 &     22 17 07.73 &     +00 19 58.3 &  1.000 & $-$8 &  1.56 &  8.51 &         44.6$^{+12.4}_{-11.2}$ &                        $<$17.0 &         42.2$^{+11.7}_{-10.4}$\\
  20 &     22 17 09.60 &     +00 18 00.1 &  1.000 & $-$8 &  1.33 &  7.24 &          31.7$^{+10.2}_{-9.0}$ &           24.1$^{+7.3}_{-6.1}$ &                        $<$23.4\\
\\
  21 &     22 17 09.82 &     +00 08 56.1 &  1.000 & $-$8 &  1.45 &  9.46 &         99.4$^{+15.6}_{-14.4}$ &          56.0$^{+10.3}_{-9.1}$ &         43.2$^{+12.4}_{-11.2}$\\
  22 &     22 17 10.04 &     +00 13 03.5 &  1.000 & $-$8 &  1.04 &  7.16 &         59.9$^{+11.7}_{-10.5}$ &           42.9$^{+8.9}_{-7.7}$ &           16.4$^{+8.4}_{-7.2}$\\
  23 &     22 17 10.10 &     +00 11 59.7 &  1.000 & $-$8 &  0.80 &  7.57 &        162.9$^{+16.4}_{-15.3}$ &        110.8$^{+12.8}_{-11.6}$ &          50.8$^{+11.1}_{-9.8}$\\
  24 &     22 17 10.42 &     +00 06 03.9 &  1.000 & $-$8 &  1.16 & 11.57 &        684.0$^{+32.0}_{-30.9}$ &        424.1$^{+23.7}_{-22.6}$ &        257.3$^{+22.2}_{-21.0}$\\
  25 &     22 17 10.60 &     +00 11 05.1 &  1.000 & $-$8 &  1.43 &  7.96 &         39.9$^{+11.7}_{-10.5}$ &           27.7$^{+7.8}_{-6.6}$ &                        $<$28.1\\
  26 &     22 17 10.77 &     +00 17 16.7 &  0.999 & $-$7 &  1.50 &  6.75 &           16.4$^{+8.2}_{-7.0}$ &                        $<$12.2 &           17.0$^{+7.8}_{-6.5}$\\
  27 &     22 17 11.13 &     +00 19 11.9 &  1.000 & $-$8 &  1.18 &  7.39 &         48.6$^{+11.2}_{-10.0}$ &           35.2$^{+8.2}_{-7.0}$ &                        $<$24.9\\
  28 &     22 17 11.26 &     +00 19 54.6 &  1.000 & $-$8 &  0.98 &  7.74 &        102.8$^{+14.1}_{-12.9}$ &          63.2$^{+10.3}_{-9.1}$ &          39.1$^{+10.5}_{-9.2}$\\
  29 &     22 17 11.94 &     +00 17 05.2 &  1.000 & $-$8 &  1.09 &  6.41 &           33.1$^{+9.3}_{-8.0}$ &           15.7$^{+6.1}_{-4.8}$ &           17.4$^{+7.7}_{-6.4}$\\
  30 &     22 17 12.04 &     +00 12 44.1 &  1.000 & $-$8 &  0.51 &  6.83 &        374.5$^{+22.7}_{-21.5}$ &           33.9$^{+8.0}_{-6.8}$ &        347.7$^{+22.0}_{-20.8}$\\
\hline
\end{tabular}
\end{center}
NOTE. --- Table~2 is presented in its entirety in the electronic version; an abbreviated version of the table is shown here for guidance as to its form and content.  The full table contains 44 columns of information for all 297 \chandra\ sources.  Meanings and units for all columns have been summarized in detail in $\S$~3.2.3.
\end{minipage}
\end{table*}

Columns~(17) and (18) give the right ascension and declination of the
near-infrared source centroid, which was obtained by matching our \hbox{X-ray}
source positions (columns~[2] and [3]) to DXS UKIDSS $K$-band positions using a
matching radius of 1.5 times the positional uncertainty quoted in column~(6).
For five \xray\ sources more than one near-infrared match was found, and for
these sources the source with the smallest offset was selected as the most
probable counterpart.  Using these criteria, 183 ($\approx$63~per~cent) of the sources
have $K$-band counterparts.  Note that the matching criterion used here is more
conservative than that used in the derivation of our positional errors
discussed above (i.e., the median value of 1.5 times the positional uncertainty
is $\approx$1.3~arcsec).  Sources with no optical counterparts have right
ascension and declination values set to \hbox{``00 00 00.00''} and \hbox{``+00
00 00.0''}.

Column~(19) indicates the measured offset between the $K$-band and \xray\
source positions in arcseconds.  Sources with no $K$-band counterparts have a
value set to ``$-$1.''  We find a median offset of 0.35~arcsec.

Column~(20) provides the corresponding $K$-band magnitude (Vega) for the source
located at the position indicated in columns~(17) and (18).  Sources with no
$K$-band counterpart have a value set to ``$-$1.''

%
%

\begin{table*}
\begin{minipage}{175mm}
\begin{center}
\caption{Summary of \chandra\ Source Detections}
\begin{tabular}{lccccccccc}
\hline\hline
 & & \multicolumn{4}{c}{Detected Counts Per Source} & \multicolumn{4}{c}{Flux Per Source}  \\
 \multicolumn{1}{c}{Band} & Number of & \multicolumn{4}{c}{\rule{2in}{0.01in}} & \multicolumn{4}{c}{\rule{2.2in}{0.01in}} \\
 \multicolumn{1}{c}{(keV)} & Sources & Maximum & Minimum & Median & Mean & Maximum$^a$ & Minimum$^a$ & Median$^a$  & Mean$^a$ \\
\hline
Full (0.5--8) \ldots \ldots \ldots \ldots  & 278 & 1965.4 & 4.8 & 33.5 & 105.4 & $-$13.2 & $-$15.8 & $-$14.7 & $-$14.3 \\
Soft (0.5--2) \ldots \ldots \ldots \ldots  & 248 & 1228.9 & 3.0 & 18.2 & 68.5 & $-$13.7 & $-$16.3 & $-$15.4 & $-$14.9 \\
Hard (2--8) \ldots \ldots \ldots \ldots    & 206 & 746.7 & 4.0 & 26.0 & 57.0 & $-$13.4 & $-$15.6 & $-$14.7 & $-$14.4 \\
\hline
\end{tabular}
\end{center}
$^{a}$Values in these columns represent the logarithm of the maximum, minimum, median, and mean fluxes in units of ergs~cm$^{-2}$~s$^{-1}$.
\end{minipage}
\end{table*}

Columns~(21)--(25) give the AB magnitudes for the Subaru $B$, $V$, $R$,
$i^\prime$, and $z^\prime$ optical bands, respectively.  Information regarding
the Subaru observations can be found in $\S$2 of Hayashino \etal\ (2004).  The
Subaru observations cover the entire \chandra\ observed region of SSA22 and
reach 5$\sigma$ limiting depths of $B =26.5$, $V=26.6$, $R=26.7$, $i^\prime =
26.4$, and $z^\prime =25.7$ AB magnitudes.  Using a constant 1.5~arcsec matching
radius, we found that 175, 202, 211, 210, and 205 of the main catalog sources
had $B$, $V$, $R$, $i^\prime$, and $z^\prime$ band counterparts, respectively;
213 of the main catalog sources had at least one optical counterpart.  Based on
the shift and rematch technique described in $\S$3.2.2, we estimate that
$\approx$45.8$^{+6.9}_{-6.7}$ are expected to be false matches.  When a
counterpart is not identified for a given IRAC band, a value of ``$-$1'' is
listed for that band.  

Columns~(26)--(29) give the AB magnitudes for the \spitzer\ IRAC bands at 3.6,
4.5, 5.8, and 8.0$\mu$m, respectively.  Information regarding the IRAC
observations can be found in $\S$2.1 of Webb \etal\ (2009).  The IRAC
observations cover the majority of the \chandra\ observed region of SSA22 and
reach 5$\sigma$ limiting depths of 23.6, 23.4, 21.6, and 21.5 AB magnitudes for
the 3.6, 4.5, 5.8, and 8.0$\mu$m bands, respectively.  Using a constant
1.5~arcsec matching radius, we found that 212, 217, 173, and 174 of the main
catalog sources had 3.6, 4.5, 5.8, and 8.0$\mu$m counterparts, respectively;
234 of the main catalog sources have at least one IRAC counterpart.  Based on
the shift and rematch technique described in $\S$3.2.2, we estimate that
$\approx$21.3$^{+5.7}_{-4.6}$ are expected to be false matches.  When a
counterpart is not identified for a given IRAC band, a value of ``$-$1'' is
listed for that band.  We note that a small area of the \chandra\ exposure has
no overlapping IRAC observations (see Fig.~2), and sources in these regions
have a value of ``$-$2'' listed in these columns.

Column~(30) provides the best available optical spectroscopic redshift for each
\xray\ source when the optical and \xray\ positions were offset by less than
1.5~arcsec.  Spectroscopic redshifts for 46 sources are provided: 31 sources from
the Garilli \etal\ (2008) catalog of the VIMOS VLT Deep Survey (VVDS; Le
F{\`e}vre \etal\ 2005), two sources from the Steidel \etal\ (2003) LBG survey,
four sources from the Matsuda \etal\ (2005) LAE survey, and nine sources from a
new spectroscopic campaign of Chapman \etal\ (2009, in-preparation).  In total,
nine of the sources had spectroscopic redshifts within \hbox{$z =$~3.06--3.12}
($\Delta v \approx$~4,000~km~s$^{-1}$), suggesting that they are likely members
of the SSA22 protocluster (see $\S$4 for further details).  Sources with no
spectroscopic redshift available have a value set to ``$-$1.''

Column~(31) indicates which optical spectroscopic survey provided the redshift
value quoted in column~(30).  Source redshifts provided by Garilli \etal\
(2008), Steidel \etal\ (2003), Matsuda \etal\ (2005), and Chapman \etal\ (2009,
in-preparation) are denoted with the integer values 1--4, respectively.
Sources with no spectroscopic redshift available have a value set to ``$-$1.''

%
%
\begin{figure*}
\centerline{
\includegraphics[width=8.5cm]{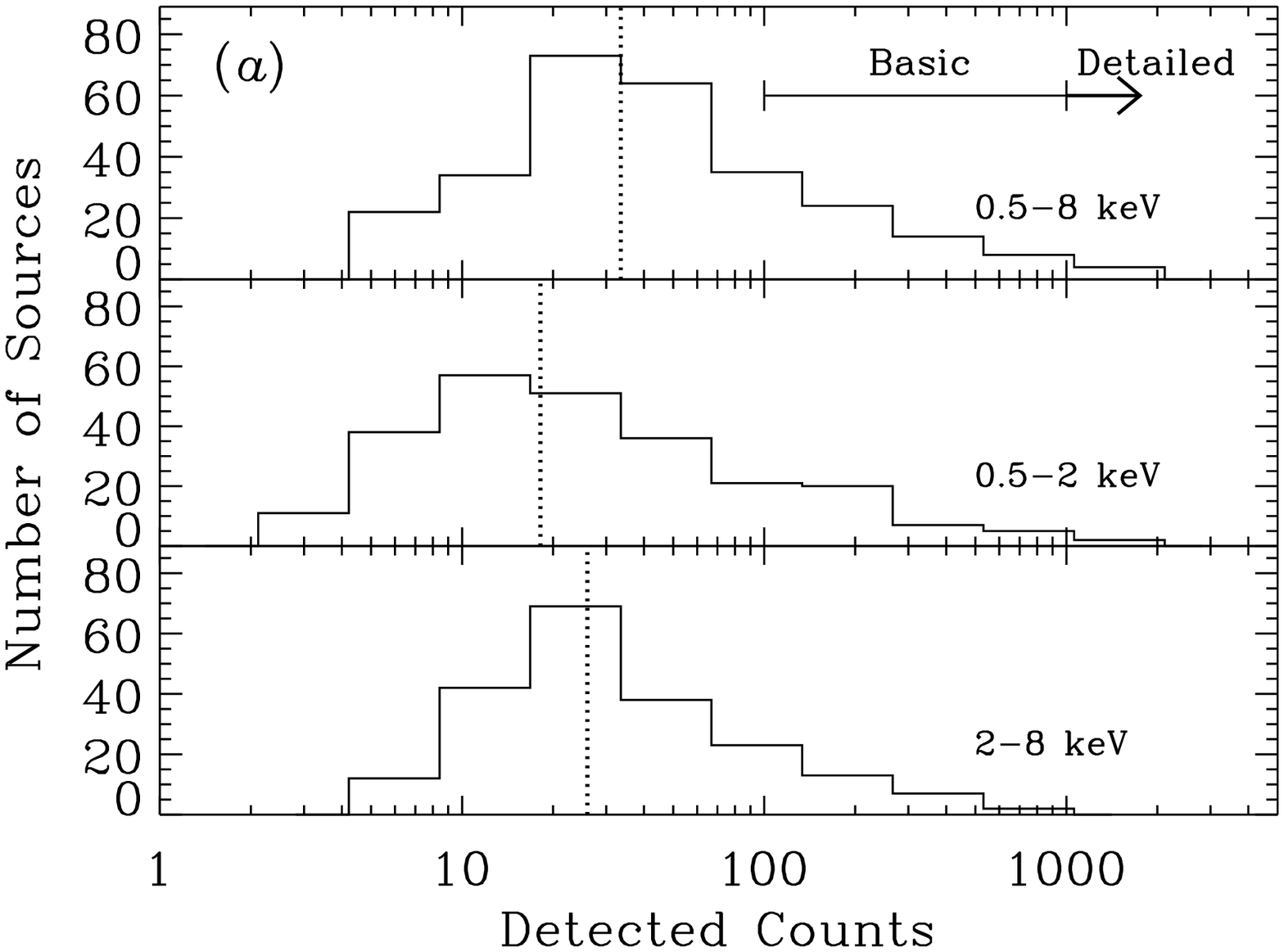}
\hfill
\includegraphics[width=8.5cm]{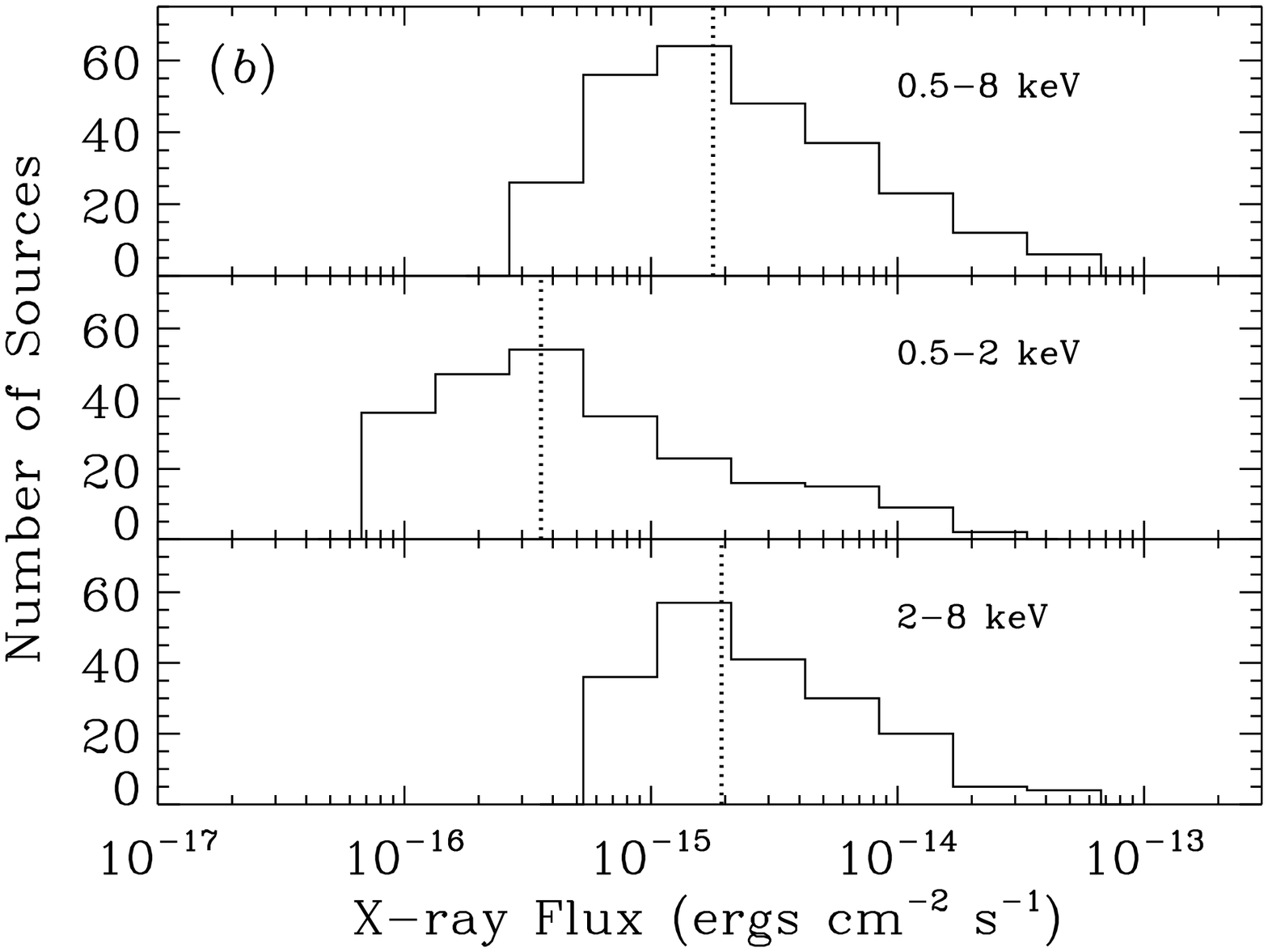}
}
\caption{
({\it a\/}) The distributions of detected source counts for
sources in the main {\it Chandra} catalog in the full ({\it top}), soft ({\it
middle}), and hard ({\it bottom}) bands.  Sources with upper limits have not
been included in this plot.  The vertical dotted lines indicate the median
number of detected counts: 32.7, 18.2, and 25.0 for the full, soft, and hard
bands, respectively.  Regions where basic (\hbox{$\approx$100--1000~counts})
and detailed ($\simgt$1,000 counts) \xray\ spectroscopic fitting is possible
are indicated.  The numbers of sources rise toward lower counts before peaking
around \hbox{$\approx$10--30~counts}, where our survey is thought to be highly
complete.
({\it b\/}) Histograms showing the distributions of \hbox{X-ray} fluxes for
sources in the main {\it Chandra} catalog in the full ({\it top}), soft ({\it
middle}), and hard ({\it bottom}) bands.  Sources with upper limits have not
been included in this plot.  The vertical dotted lines indicate the median
fluxes of 15.8, 4.0, and \hbox{20.0~$\times$~10$^{-16}$ \flux} for the full,
soft, and hard bands, respectively.
}
\end{figure*}

Columns~(32)--(34) give the effective exposure times derived from the
standard-band exposure maps (see \S3.1 for details on the exposure maps).
Dividing the counts listed in columns~(8)--(16) by the corresponding effective
exposures will provide vignetting-corrected and quantum efficiency
degradation-corrected count rates.

Columns~(35)--(37) give the band ratio, defined as the ratio of counts between
the hard and soft bands, and the corresponding upper and lower errors,
respectively. Quoted band ratios have been corrected for differential
vignetting between the hard band and soft band using the appropriate exposure
maps. Errors for this quantity are calculated following the numerical error
propagation method described in \S1.7.3 of Lyons (1991); this avoids the
failure of the standard approximate variance formula when the number of counts
is small (see \S2.4.5 of Eadie et~al. 1971) and has an error distribution that
is non-Gaussian.  Upper limits are calculated for sources detected in the soft
band but not the hard band and lower limits are calculated for sources detected
in the hard band but not the soft band.  For these sources, the upper and lower
errors are set to the computed band ratio.  Sources detected only in the full
band have band ratios and corresponding errors set to ``$-1$.''

Columns~(38)--(40) give the effective photon index ($\Gamma_{\rm eff}$) with
upper and lower errors, respectively, for a power-law model with the Galactic
column density. The effective photon index has been calculated based on the
band ratio in column~(35).  

For sources that are not detected (as per the definition discussed above in the
description of columns~[8]--[16]) in the hard band or soft band, then lower or
upper limits, respectively, are placed on $\Gamma_{\rm eff}$; in these cases,
the upper and lower errors are set to the limit that is provided in
column~(38).  When a source is only detected in the full band, then the
effective photon index and upper and lower limits are set to 1.4, a value that
is representative for faint sources that should give reasonable fluxes.

Columns~(41)--(43) give observed-frame fluxes in the three standard bands;
quoted fluxes are in units of ergs~cm$^{-2}$~s$^{-1}$.
Fluxes have been computed using the counts in \hbox{columns~(8), (11), and
(14)}, the appropriate exposure maps (columns~[32]--[34]), and the spectral
slopes given in column~(38). Negative flux values indicate upper limits.  The
fluxes have been corrected for absorption by the Galaxy but have not been
corrected for material intrinsic to the source.  For a power-law model with
$\Gamma=1.4$, the soft-band and hard-band Galactic absorption corrections are
$\approx$12.6~per~cent and $\approx$0.4~per~cent, respectively.  We note that, due to the
Eddington bias, sources with a low number of net counts ($\simlt$10 counts) may
have true fluxes lower than those computed using the basic method used here
(see, e.g., Vikhlinin \etal\ 1995; Georgakakis \etal\ 2008).  However, we aim
to provide only observed fluxes here and do not make corrections for the
Eddington bias.  More accurate fluxes for these sources would require (1) the
use of a number-count distribution prior to estimate the flux probabilities for
sources near the sensitivity limit and (2) the direct fitting of the \xray\
spectra for each observation; these analyses are beyond the scope of the
present paper.

Column~(44) gives notes on the sources.
``O'' refers to objects that have large cross-band (i.e., between the three
standard bands) positional offsets ($>2$~arcsec); all of these sources lie at
off-axis angles $>$5.5~arcmin.
``S'' refers to close-double or close-triple sources where manual separation
was required (see discussion in $\S$3.2.1).
``D'' refers to a source having an obvious diffraction spike in the $K$-band
image, suggesting the source is likely a Galactic star.
``LBG'' and ``LAE'' indicate sources included in the Steidel \etal\ (2003) LBG
survey and the Hayashino \etal\ (2004) LAE survey, respectively.
``LAB'' indicates that the source was coincident with a LAB in the Geach \etal\
(2009) study.

%
%

\begin{table}
\begin{center}
\caption{Sources Detected in One Band but not Another}
\begin{tabular}{lccc}
\hline\hline
  &  \multicolumn{3}{c}{Nondetected Energy Band}  \\
 \multicolumn{1}{c}{Detection Band} &  \multicolumn{3}{c}{\rule{1.1in}{0.01in}}  \\
 \multicolumn{1}{c}{(keV)} & Full & Soft & Hard  \\
\hline
Full Band (0.5--8) \ldots\ldots\ldots\ldots\ldots &  \ldots & 46 & 75 \\
Soft Band (0.5--2) \ldots\ldots\ldots\ldots\ldots & 16 & \ldots & 82 \\
Hard Band (2--8) \ldots\ldots\ldots\ldots\ldots\ldots & 3 & 40 & \ldots \\
\hline
\end{tabular}
\end{center}
NOTE. --- For example, of the 278 full band detected sources, there were 46 sources detected in the full band that were not detected in the soft band.
\end{table}

%
%

\begin{table*}
\begin{minipage}{175mm}
\begin{center}
\caption{Background Parameters}
\begin{tabular}{lcccc}
\hline\hline
  &  \multicolumn{2}{c}{Mean Background} & & \\
 \multicolumn{1}{c}{Bandpass} & \multicolumn{2}{c}{\rule{2in}{0.01in}} & Total Background$^c$ & Count Ratio$^d$  \\
 \multicolumn{1}{c}{(keV)} & (counts pixel$^{-1}$)$^a$ & (counts pixel$^{-1}$ Ms$^{-1}$)$^b$ & ($10^5$ counts) & (background/source) \\
\hline
Full Band (0.5--8) \ldots\ldots\ldots\ldots\ldots\ldots\ldots & 0.072 & 0.267 & 3.6 & 12.1 \\
Soft Band (0.5--2) \ldots\ldots\ldots\ldots\ldots\ldots\ldots & 0.017 & 0.065 & 0.9 & 5.0 \\
Hard Band (2--8) \ldots\ldots\ldots\ldots\ldots\ldots\ldots\ldots & 0.055 & 0.191 & 2.7 & 23.0 \\
\hline
\end{tabular}
\end{center}
$^{a}$The mean numbers of background counts per pixel. These are measured from the background images described in $\S$~3.3.\\
$^{b}$The mean numbers of counts per pixel divided by the mean effective exposure. These are measured from the exposure maps and background images described in $\S$~3.3.\\
$^{c}$Total number of background counts in units of $10^5$ counts.\\
$^{d}$Ratio of the total number of background counts to the total number of net source counts.
\end{minipage}
\end{table*}

In Table~3 we summarise the source detections, counts, and fluxes for the
three standard bands for the main {\it Chandra} catalog.  In total 297 point
sources are detected (i.e., with \ae-computed binomial probabilities of $P <
0.01$) in one or more of the three standard bands with 278, 248, and 206
sources detected in the full, soft, and hard bands, respectively.  In Table~4
we summarise the number of sources detected in one band but not another.  All
but three of the sources are detected in either the soft or full bands, which
is similar to that found for the $\approx$2~Ms \cdfn\ (one source) and
$\approx$2~Ms \cdfs\ (three sources).  From Tables~3 and 4, we find that the
fraction of hard-band sources not detected in the soft band is $40/206 \approx
19$~per~cent, similar to that found in the $\approx$250~ks \ecdfs, yet somewhat larger
than that found in the $\approx$2~Ms \cdfn\ and $\approx$2~Ms \cdfs, where the
fraction is $\approx$14~per~cent.  In Figure~7$a$, we show the distributions of
detected counts in the three standard bands.  There are 60 sources with $>$100
full-band counts, for which spectral analyses are possible; there are four
sources with $>$1000 full-band counts.  We note that the number of sources
continuously rises with decreasing detected counts before peaking around the
completeness limit of our survey, which occurs at \hbox{$\approx$10--30~counts}
depending on the bandpass.  In Figure~7$b$ we show the distributions of
\hbox{X-ray} flux in the three standard bands.  The \hbox{X-ray} fluxes in this
survey span roughly three orders of magnitude and have median fluxes of 15.8,
4.0, and $20.0 \times 10^{-16}$~\flux\ in the full, soft, and hard bands,
respectively.

%
\subsection{Background and Sensitivity Analysis}
%

The faintest sources in our main catalog have \hbox{$\approx$3} counts (see
Table~3).  Assuming a $\Gamma=$~1.4 power law \xray\ spectrum with Galactic
absorption as given in $\S$1, the corresponding soft-band and hard-band fluxes
at the aim points are $\approx$5.0~$\times$~$10^{-17}$ \flux\ and
$\approx$2.4~$\times$~$10^{-16}$ \flux, respectively. This gives a measure of
the ultimate sensitivity of the SSA22 survey; however, these values are only
relevant for a small region near the aim point and are also subject to
significant incompleteness due to Poisson fluctuations of source and background
counts at these levels. To determine the sensitivity as a function of position
within the SSA22 field, it is necessary to account for the broadening of the PSF
with off-axis angle and changes in the effective exposure (due to, e.g.,
vignetting and chip gaps; see Fig.~3$c$) and background rate across the field.
We estimated the sensitivity across the field by calibrating the relationship
between the total number of extracted source counts $S$ versus the local
background counts $B$ for sources detected in our main catalog.  This was
achieved mathematically through the use of a binomial probability model, which
estimates the value of $S$ given $B$ background counts when a binomial
probability of $P = 0.01$ is required for a detection.  Our resulting relation
is

\begin{equation}
\log(S)~=\alpha + \beta \log(B) + \gamma [\log(B)]^2 + \delta [\log(B)]^3
\end{equation}

\noindent where $\alpha = 0.6832$, $\beta = 0.4956$, $\gamma = 0.1124$, and
$\delta = 0.0003$ are fitting constants.  We note that this equation has the
same functional form as that used by Lehmer \etal\ (2005) and Luo \etal\
(2008), which is appropriate for sources detected using {\ttfamily wavdetect}
at a false-positive probability threshold of $10^{-6}$.  However, the constant
values differ mildly (most notably in the value of $\alpha$) from those used by
Lehmer \etal\ (2005) and Luo \etal\ (2008), due to the different detection
criteria adopted in this paper.  In equation~2, the only component that we need
to measure is the local background $B$.  To be consistent with our adopted
detection criteria described in $\S\S$~3.2.1 and 3.2.2, we measured the local
background in a source cell using the background maps described below assuming
an aperture size of 90~per~cent of the encircled-energy fraction of the PSF; for ease
of computation we utilised circular extraction apertures when measuring local
backgrounds (see footnote~2 for circular approximations to the 90~per~cent encircled
energy fraction).  The total background includes contributions from the
unresolved cosmic \xray\ background, particle background, and instrumental
background (e.g.,\ Markevitch 2001; Markevitch \etal\ 2003; Worsley \etal\
2005; Hickox \& Markevitch 2006), and for our analyses, we are interested in
the total background and therefore do not distinguish between these different
components.

We created background maps for all of the three standard-band images by first
masking out point sources from our main catalog using circular apertures with
radii of $1.1 \times$ the $\approx$99.9~per~cent PSF encircled-energy fraction radii
as defined in $\S$3.2.1.  As a result of this masking procedure, the background
maps include minimal contributions from main catalog point sources.  They will,
however, include \xray\ counts from the extended sources (e.g., the source
\hbox{J221744.6+001738} described in $\S$5 below), which will cause a mild
overestimation of the measured background near and within this source.
Extensive testing of the background count distributions in all three standard
bandpasses has shown that the \xray\ background is nearly Poissonian (see
$\S$4.2 of Alexander \etal\ 2003).  We therefore filled in the masked regions
for each source with local background counts that were estimated using the
probability distribution of counts within an annulus with an inner radius equal
to that of the masked out region (i.e., the $1.1 \times$ the $\approx$99.9~per~cent
PSF encircled-energy fraction radius) and an outer radius equal to the size of
the background extraction radius defined in $\S$3.2.1; here, the outer radii
have sizes in the range of \hbox{$\approx$1.5--3.7} times the inner radii.  The
background properties are provided in Table~5.  The majority of the pixels have
no background counts (e.g., in the full band $\approx$93~per~cent of the pixels are
zero) and the mean background count rates for these observations are broadly
consistent with those presented in Alexander \etal\ (2003) and Luo \etal\
(2008).

Following equation 2, we created sensitivity maps using the background and
exposure maps.  We assumed a $\Gamma=$~1.4 power-law \xray\ spectral energy
distribution with Galactic absorption.  In Figure~3$d$ we show the full-band
sensitivity map, and in Figure~4$b$ we plot the flux limit versus solid angle
for the full, soft, and hard bands.  We note that the most sensitive
$\approx$10~arcmin$^2$ region near the aim point has average \hbox{0.5--2~keV}
and \hbox{2--8~keV} sensitivity limits of \hbox{$\approx$$5.7 \times
10^{-17}$~ergs~cm$^{-2}$~s$^{-1}$} and \hbox{$\approx$$3.0 \times
10^{-16}$~ergs~cm$^{-2}$~s$^{-1}$}, respectively.  

%
%
\begin{figure}
\centerline{
\includegraphics[width=9.5cm]{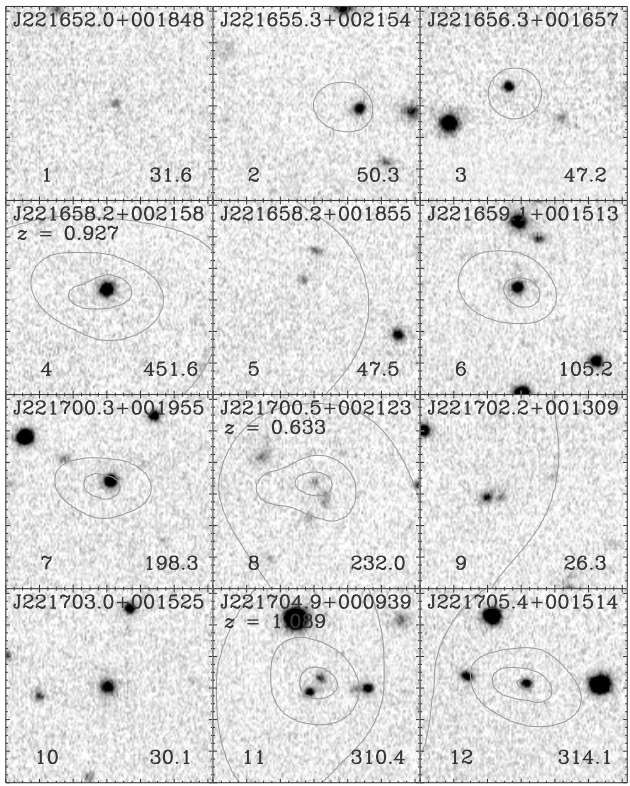}
}
\caption{
DXS $K$-band postage-stamp images for the sources in the main \chandra\ catalog
with adaptively smoothed \xray\ contours overlaid.  The contours are
logarithmic in scale and range from $\approx$0.00032--32~per~cent (with seven
total logarithmic scales) of the maximum pixel value in the whole
$\approx$330~arcmin$^2$ image.  The source name, composed of the sexagesimal
J2000 source coordinates, has been labelled at the top of each image.  The
source number (column~[1] of Table~2) and full-band net source counts
(column~[8] of Table~2) have been provided in the lower-left and lower-right
corners, respectively; when available, the value of the spectroscopic redshift
has been indicated below the source name.  Each image is $\approx$24.6~arcsec
on a side, and the source of interest is always located at the centre of the
image.  In several cases no \xray\ contours are present, either because these
sources were not detected in the full band or the full-band counts are low and
{\ttfamily CSMOOTH} has suppressed the observable emission in the adaptively
smoothed image.  Only one of the 24 pages of cutouts is included here; all 24
pages are available at the \hbox{SSA22} website.
}
\end{figure}

%
%
\begin{figure}
\centerline{
\includegraphics[width=9cm]{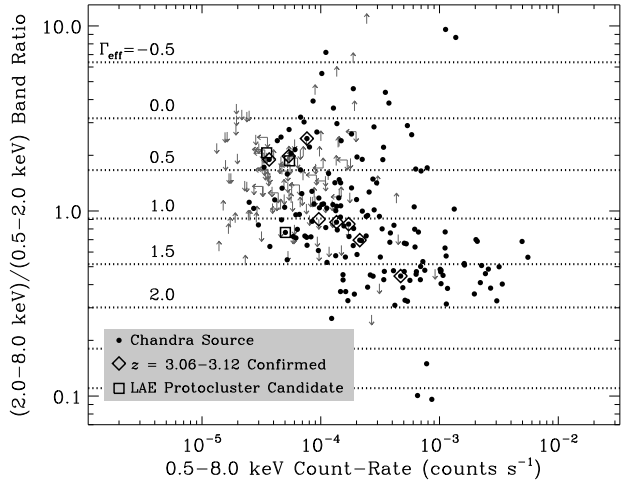}
}
\caption{
Hard-band--to--soft-band count-rate ratio (i.e.,
$\Phi_{\rm 2-8~keV}/\Phi_{\rm 0.5-2~keV}$, where $\Phi$ is the count-rate in
units of counts~s$^{-1}$) versus full-band count-rate for sources in the main
\chandra\ catalog.  We note that sources that were only detected in the full-band
cannot be plotted in this figure as their band ratios are unconstrained.
Horizontal dotted lines show the band ratios corresponding to given effective
photon indices; these were calculated using PIMMS.  Sources that have been
spectroscopically confirmed to lie in the SSA22 protocluster (\hbox{$z
=$~3.06--3.12}) have been indicated with open diamonds and additional LAEs that
are likely members of the protocluster are shown with open squares.  We note
that one of the protocluster sources has not been plotted here, since it was
only detected in the full band. 
}
\end{figure}

%
\section{Multiwavelength Properties of Main Catalog Sources}
%

In this section, we utilise the multiwavelength data in SSA22 to explore the
range of source types detected in our main \chandra\ catalog and highlight
the basic properties of the sources in the protocluster at $z = 3.09$.  In
Figure~8 we show ``postage-stamp'' images from the DXS $K$-band image with
adaptively-smoothed full band contours overlaid for sources included in the
main \chandra\ catalog.  The wide range of \xray\ source sizes observed in
these images is largely due to PSF broadening with off-axis angle.  The
postage-stamp images show a wide variety of $K$-band source types including
unresolved point sources, bright Galactic (Milky Way) stars, extended galaxies,
and sources without any obvious counterpart (see column~[44] in Table~2).

In Figure~9 we show the band ratio as a function of full-band count rate for
sources in the main \chandra\ catalog.  This plot shows that the mean band
ratio for sources detected in both the soft and hard bands hardens for fainter
fluxes, a trend also observed in the CDFs (e.g., Alexander \etal\ 2003; Lehmer
\etal\ 2005; Tozzi \etal\ 2006; Luo \etal\ 2008).  This trend is due to the
detection of more absorbed AGNs at low flux levels, and it has been shown that
AGNs will continue to dominate the number counts down to \hbox{0.5--2~keV}
fluxes of $\approx$(2--6)~$\times$~10$^{-18}$~\flux\ (e.g., Bauer \etal\ 2004;
Kim \etal\ 2006); however, we expect that an important minority of the sources
detected near the flux limit ($\simlt$$3 \times 10^{-16}$~\flux\ in the
\hbox{0.5--2~keV} band) are normal galaxies (e.g., Lehmer \etal\ 2007, 2008;
Ptak \etal\ 2007).  As mentioned above, nine sources in our main catalog have
spectroscopic redshifts of \hbox{$z =$~3.06--3.12}, suggesting that they are
members of the SSA22 protocluster.  If we include LAEs selected to lie at
$z\approx 3.1$ as being likely protocluster members, then we have a total of 12
\xray\ detected protocluster sources.  From Figure~9, we see that these
protocluster sources occupy a range of band-ratios and full band count rates
similar to all main catalog sources and do not appear to reside in any unique
region of parameter space.

%
%
\begin{figure*}
\centerline{
\includegraphics[width=8.5cm]{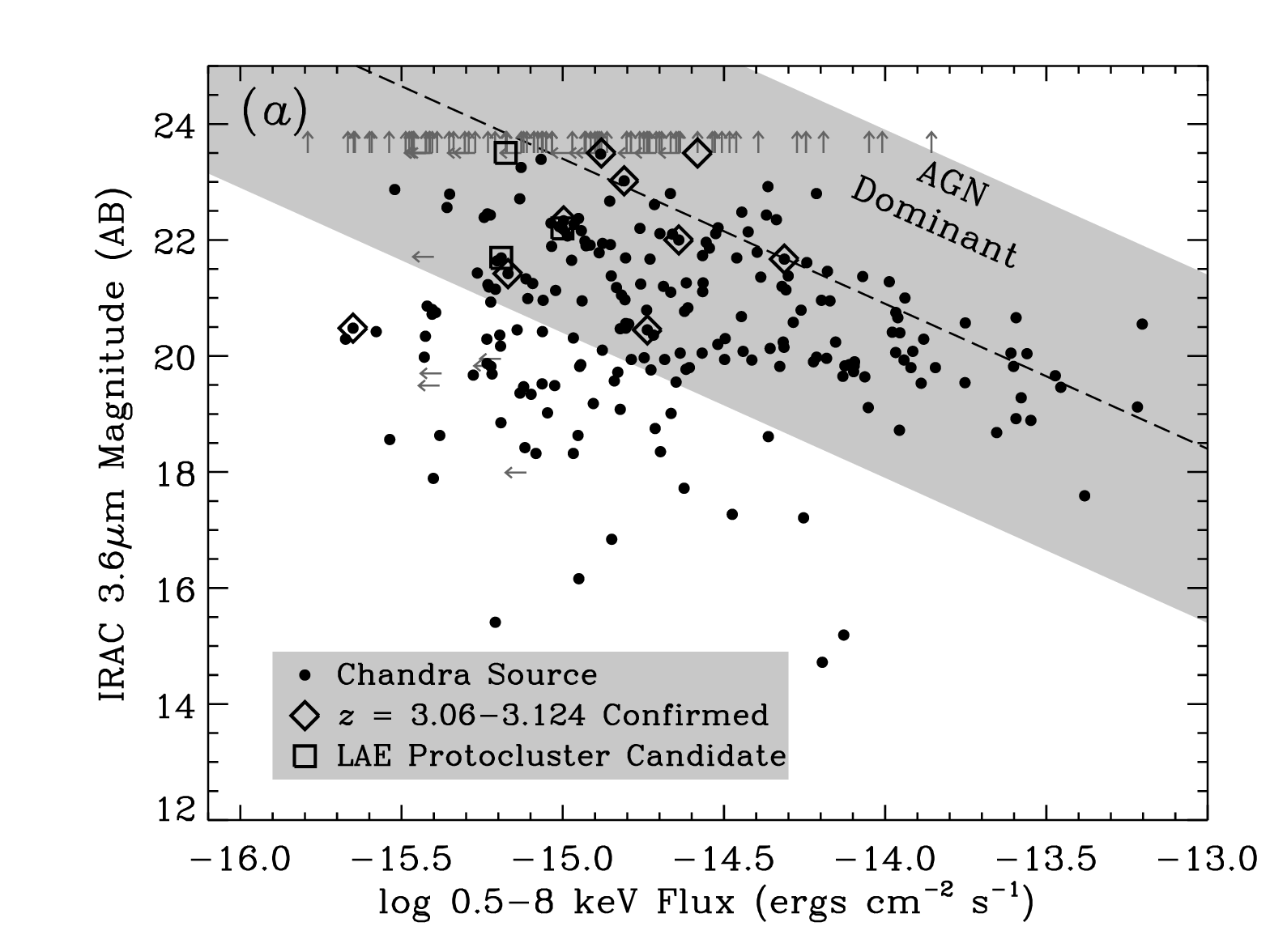}
\hfill
\includegraphics[width=8.5cm]{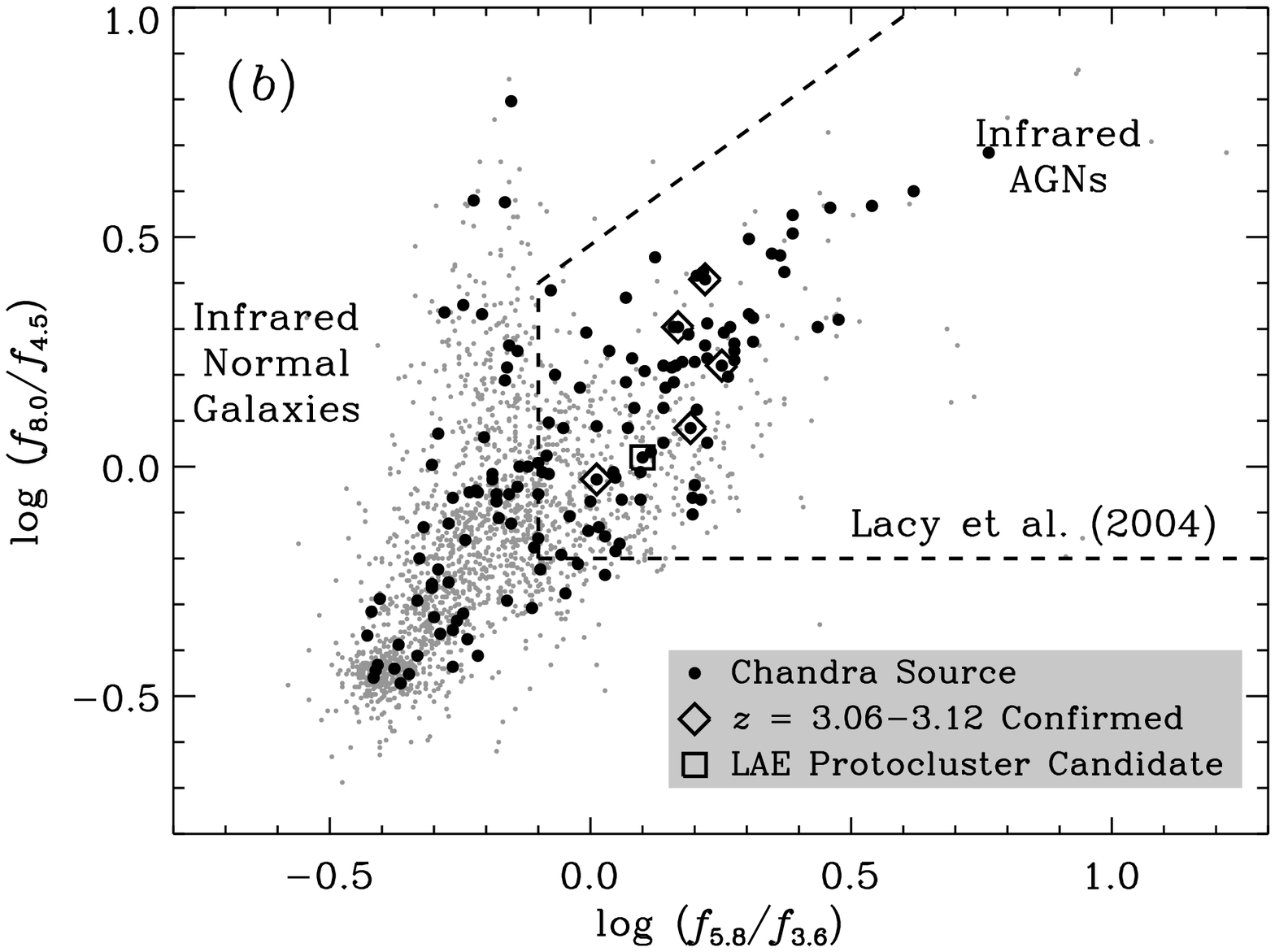}
}
\caption{
({\it a\/}) Apparent IRAC 3.6$\mu$m magnitude (AB) versus the logarithm of the
\hbox{0.5--8~keV} flux for sources in the main \chandra\ catalog.  In total 212
sources ($\approx$71~per~cent) have 3.6$\mu$m counterparts down to a limiting
magnitude of $\approx$23.5~mag.  The dashed line and shaded envelope and
represents the region where AGN activity is expected to dominate both the
3.6$\mu$m and \xray\ emission (see discussion in $\S$4 for details).  The
majority of our sources have 3.6$\mu$m and full band fluxes that place them in
the shaded region, indicating that many of these sources are likely to be AGNs.
A significant minority of sources below full band fluxes of $\approx$$3 \times
10^{-15}$~\flux\ have relatively low 3.6$\mu$m to \xray\ flux ratios.  These
sources are likely to be a mixture of obscured or low-luminosity AGNs,
normal/starburst galaxies, and Galactic stars.
({\it b\/}) Logarithm of the 8.0$\mu$m to 4.5$\mu$m flux density ratio versus
the logarithm of the 5.8$\mu$m to 3.6$\mu$m flux density ratio for sources in
the SSA22 field that are detected in all four IRAC bands.  We have shown all
SSA22 sources from the full IRAC catalog ({\it small gray circles\/}) and all
\chandra\ detected sources with IRAC counterparts ({\it black circles\/}).  The
Lacy \etal\ (2004) colour-criterion region used to identify AGN candidates has
been highlighted with a dashed line boundary.  
}
\end{figure*}

Figure~10$a$ shows the IRAC 3.6$\mu$m magnitude (AB) versus the full-band flux for
sources included in the main catalog.  The approximate \hbox{X-ray} to
3.6$\mu$m flux ratio range expected for AGN-dominated systems is indicated.
This ``AGN region'' was calibrated using the 28 \xray-detected broad-line
quasars studied by Richards \etal\ (2006) and represents the mean logarithm of
the 3.6$\mu$m to \xray\ flux ration and its 3$\sigma$ scatter.  We note that
this categorization is only appropriate for powerful AGNs where both \xray\ and
infrared emission is likely to be dominated by the AGN component (i.e., with
little fractional contribution from galactic emission).  Therefore, heavily
obscured or low-luminosity AGNs that are faint in the \xray\ band, but bright
in the infrared, can be pushed out of the marked AGN region and into the realm
where normal/starburst galaxies and Galactic stars are expected to be found.
We find that the majority of our sources lie in the designated AGN region;
however, a significant minority of the sources appear to have small \xray\ to
3.6$\mu$m flux ratios; these sources are either obscured or low-luminosity
AGNs, normal/starburst galaxies, or Galactic stars.  The majority of the
confirmed protocluster sources are found in the AGN region of Figure~10$a$ (i.e.,
the {\it shaded region\/}) with the exception of \hbox{J221742.0+001913}, which
is \xray\ faint but has IRAC colours (see, e.g., Lacy \etal\ 2004 and Fig.~10$b$)
characteristic of AGNs, and it is therefore likely to be powered by a
heavily-obscured AGN.

%
%
\begin{figure}
\centerline{
\includegraphics[width=9.5cm]{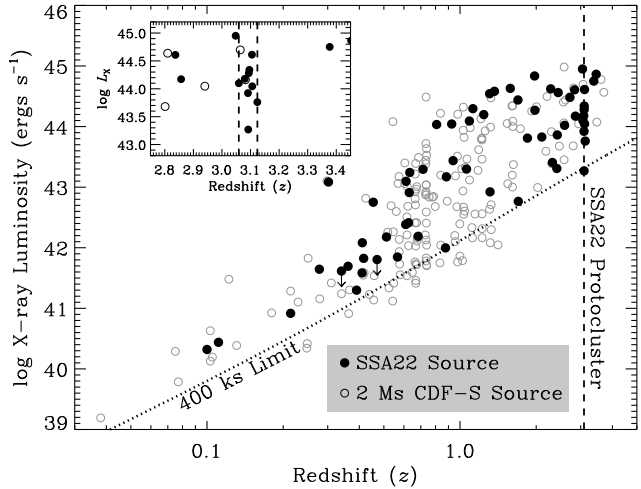}
}
\caption{
Logarithm of the observed-frame 0.5--8~keV luminosity (i.e., the rest-frame
[0.5--8~keV]~$\times$~[$1+z$] band) versus redshift for sources in our sample
that have optical spectroscopic redshifts (see column~[30] of Table~2).  The
dotted curve shows the average luminosity limit at 5~arcmin off-axis.  This
limit was determined using the \hbox{0.5--8~keV} sensitivity map (see $\S$3.3)
and our adopted cosmology (see $\S$1).  For comparison, we have plotted the
sources from the $\approx$2~Ms \cdfs\ ({\it open circles\/}; Luo \etal\ 2008).
The vertical dashed line highlights the location of the SSA22 protocluster at
$z = 3.09$, and in the inset plot, we show the region of the plot around the
protocluster.  This shows the relatively large number of protocluster sources
compared to what is found in a low density field like the \cdfs.  There are two
\cdfs\ sources within the redshift boundaries of the protocluster; however,
only one can be seen due to the large number of SSA22 protocluster sources in
this parameter space.
}
\end{figure}

In Figure~10$b$, we show the 8.0$\mu$m to 4.5$\mu$m versus 5.8$\mu$m to
3.6$\mu$m IRAC flux density ratios for sources in the SSA22 IRAC catalogs that
are detected in all four IRAC bandpasses (see columns~[26]--[29] in Table~2).
The AGN region of this diagram, as defined by Lacy \etal\ (2004), has been
outlined with dashed lines.  Of the 147 main catalog sources detected in all
four IRAC bands, we find that 81 ($\approx$55~per~cent) of them lie in the AGN
region in Figure~10$b$, as compared with $\approx$25~per~cent of all sources in
the full IRAC catalog.  This result is consistent with the fact that the
majority of the \xray\ detected sources are likely AGNs.  It is of interest to
note that a significant number (66; $\approx$45~per~cent) of the \xray-detected
sources are not classified as AGNs by the infrared data.  While some of these
\xray\ sources are likely to be normal/starburst galaxies and Galactic stars at
least 15 ($\approx$23~per~cent) have \xray\ properties characteristic of
obscured AGNs (i.e., $\Gamma_{\rm eff} \simlt 1$; see, e.g., Donley \etal\ 2008
for a more detailed study of similar source types in the \cdfs).  We note that
all of the six protocluster sources (i.e., those with spectroscopic redshifts
\hbox{$z =$~3.06--3.12} or LAEs) with detections in all four IRAC bands lie in
the AGN region in Figure~10$b$.

%
%
\begin{figure*}
\centerline{
\includegraphics[width=9cm]{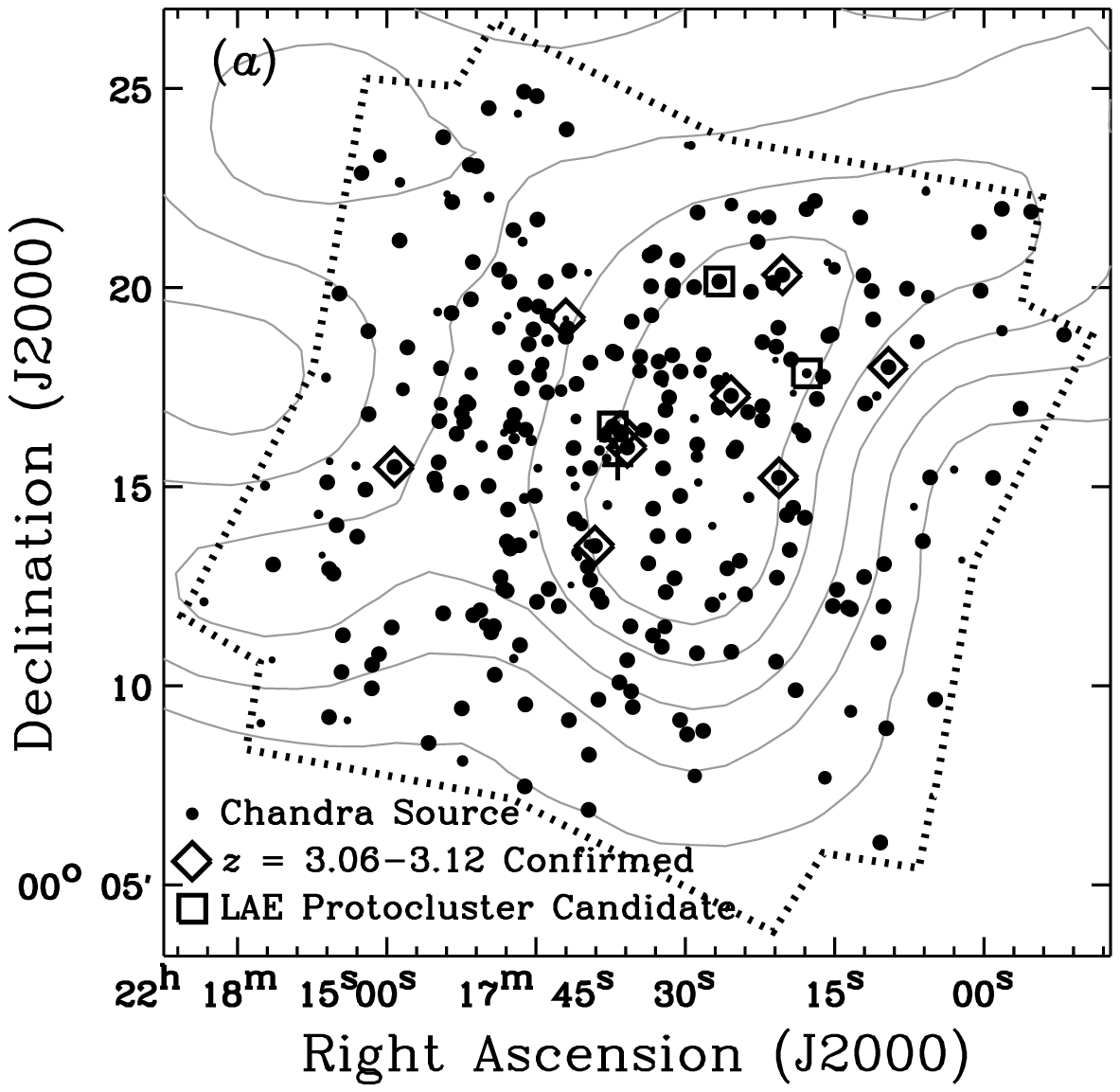}
\hfill
\includegraphics[width=9cm]{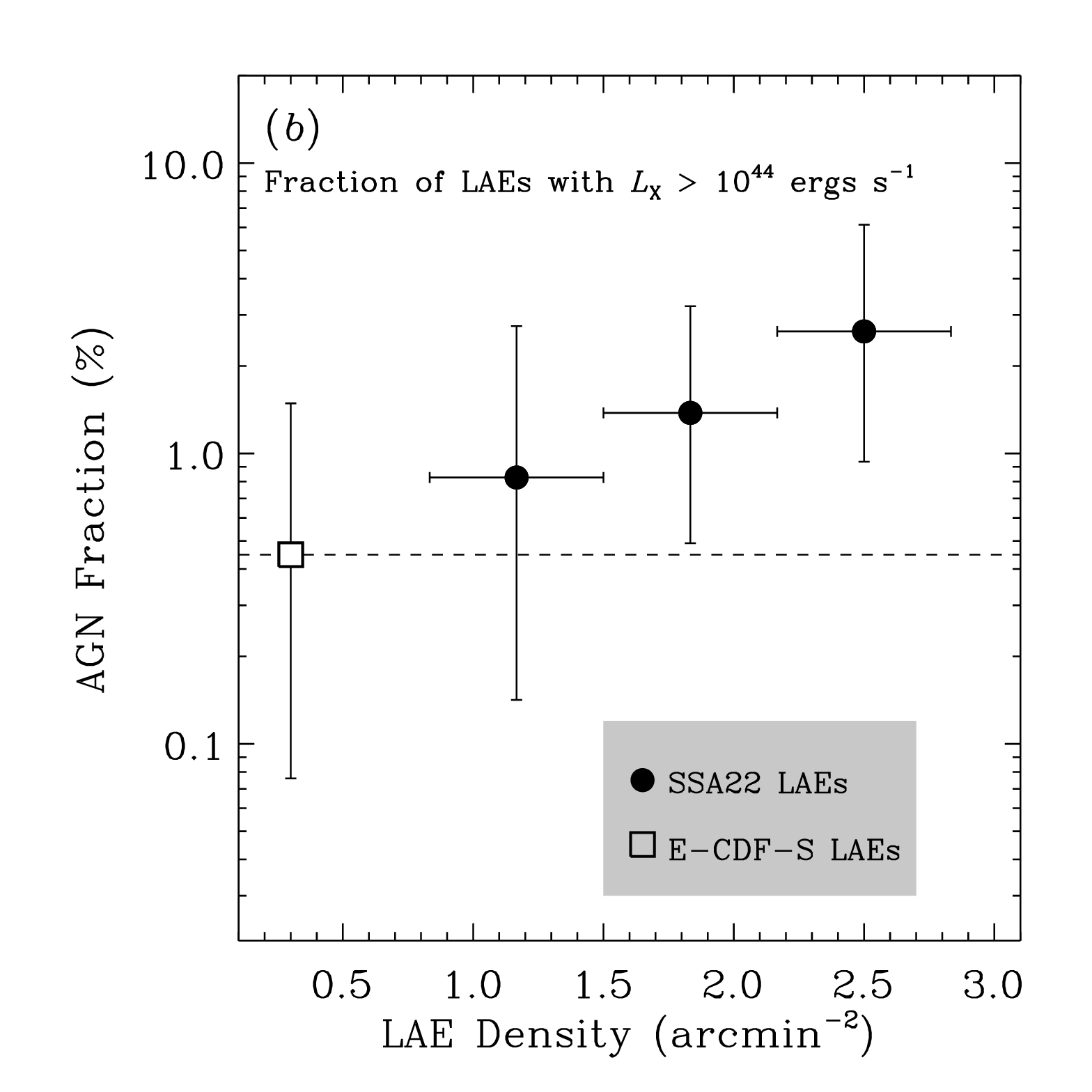}
}
\caption{
({\it a\/}) Positions of main \chandra\ catalog sources.  The dotted polygonal
and gray contours indicate the \chandra\ observed region and $z \approx 3.1$
LAE source density levels, respectively (as per Fig.~2).  All \xray--detected
sources have been indicated with small filled circles with symbol sizes
dependent on \ae\ detection significance, such that more significant sources
have larger circles; the largest symbols represent sources with $P < 10^{-5}$.
We find that the majority of the \xray\ detected protocluster sources lie in
the highest LAE density regions.
({\it b\/}) Fraction of $z \approx 3$ LAEs hosting an AGN with rest-frame
\hbox{8--32~keV} luminosity larger than $10^{44}$~\xlum\ ({\it filled
circles\/}) as a function of local LAE density.  The field-averaged AGN
fraction for $z \approx 3$ LAEs in the \ecdfs\ (from Gronwall \etal\ 2007) has
been indicated with an open square and a dashed horizontal line for guidance.
These data show suggestive evidence that the $z \approx 3$ LAE AGN fraction
increases with increasing LAE density.  In the highest density environments
(2.5~LAEs~per~arcmin$^2$) the AGN fraction is $\approx$6 times larger than in
the field ($\approx$0.3~LAEs~per~arcmin$^2$).
}
\end{figure*}

In Figure~11, we show the observed-frame full-band luminosity versus redshift
for sources in our sample and compare it with sources given in the
$\approx$2~Ms \cdfs\ sample (from Luo \etal\ 2008).  Luminosities were
calculated for sources with secure optical spectroscopic redshifts (column~[30]
of Table~2) using the full band flux provided in column~(41) of Table~2 and our
adopted cosmology (see $\S$1).
For the most part, the main catalog sources having spectroscopic redshifts span
a similar range of \xray\ luminosities and redshifts as those detected in the
$\approx$2~Ms \cdfs, with the exception of the large clustering of SSA22
sources in the $z = 3.09$ protocluster.  In the inset plot, we have highlighted
the redshift range near the protocluster, which shows the nine main catalog
sources in the protocluster (\hbox{$z =$~3.06--3.12}) compared with the two
found in this redshift range in the \cdfs.  We note that one of these sources
\hbox{J221720.6+001513} was not previously reported by Lehmer \etal\ (2009) and
was only recently identified via the new spectroscopic campaign by
Chapman \etal\ (2009, in-preparation).

In Figure~12$a$, we plot the positions of sources detected in our main catalog.
We have highlighted sources that are likely associated with the protocluster
(i.e., those with spectroscopic redshifts \hbox{$z =$~3.06--3.12} or LAEs).  We
note that the majority of these sources lie in the highest LAE density regions
due to a combination of (1) the presence of larger numbers of $z = 3.09$
objects, (2) the \xray\ imaging being most sensitive in these regions, and (3)
the possibility that the AGN fraction increases with increasing LAE density.

In Lehmer \etal\ (2009), we found that the fraction of LAEs hosting an AGN was
a factor of $\approx$6 times larger in the SSA22 protocluster than in the field
(the \ecdfs).  Using the spatial densities of LAEs in SSA22 (illustrated in
Figure~12$a$) and the methods outlined in $\S$4.2 of Lehmer \etal\ (2009), we
have now constrained how the LAE AGN fraction varies as a function of local LAE
density.  These methods account for the spatially varying \xray\ sensitivity of
the \chandra\ imaging using the sensitivity maps constructed in $\S$3.3.  In
Figure~12$b$, we show the fraction of LAEs hosting an AGN with rest-frame
\hbox{8--32~keV} luminosity greater than $10^{44}$~\xlum\ versus local LAE
density (computed as the number of LAEs per 3~arcmin radius circle) for the
479 LAEs presented in Hayashino \etal\ (2004) that fell within
\chandra-observed regions of SSA22.  For comparison, we have also included the
corresponding AGN fraction for 257 LAEs in the \ecdfs\ drawn from the Gronwall
\etal\ (2007) sample (see Lehmer \etal\ 2009 for details).  We find evidence
suggesting that the AGN activity per galaxy increases with local LAE density.
In the lowest density regions ($\approx$0.3~LAEs arcmin$^{-2}$) the AGN
fraction is $\approx$0.5~per~cent, as compared with $\approx$3~per~cent in the highest
density regions ($\approx$2.5~LAEs arcmin$^{-2}$). Using the four data points
provided in Figure~12$b$ and the Kendall's $\tau$ rank correlation statistic,
we found that the AGN fraction is positively correlated with LAE density at the
$\approx$96~per~cent confidence level; however, these data can be well fit by a
constant average AGN fraction of $\approx$1.3~per~cent ($\chi^2 = 2.0$ for 3 degrees
of freedom).  Due to small number statistics, this result is only suggestive at
present.  A more complete census of the $z = 3.1$ galaxy population in SSA22
(e.g., through wider LBG selection and spectroscopic follow-up than that
available) as well as observations of similar $z \approx 3$ structures will
improve these constraints.

%
\section{The Extended X-ray Source J221744.6+001738}
%

Through visual inspection of the adaptively smoothed images discussed in
$\S$3.1, we identified one obvious extended \xray\ source
\hbox{J221744.6+001738}.  The soft emission from this source is clearly visible
as a ``glow'' just north-west of the average aim point in Figures~3$a$ and
3$b$.

Using the adaptively-smoothed soft-band image, we defined an elliptical
aperture from which to extract \xray\ properties for the extended source.  The
elliptical aperture closely matches the apparent extent of the \xray\ emission
that is $\simgt$10~per~cent above the background level; the aperture has a
semi-major axis of 47.9~arcsec, a semi-minor axis of 27.1~arcsec, and a
position angle of 177.6~degrees clockwise from north.  Source counts $s_{\rm
ext}$ were extracted using manual aperture photometry from within the
elliptical aperture; in this process, point sources from the main catalog
(presented in Table~2) were masked out using circular apertures with radii of
1.1$\times$ the $\approx$99.9~per~cent PSF encircled energy fraction.  The
local background was estimated using an elliptical annulus with inner and outer
sizes of 1.5 and 2.5 times those used for extracting source counts.  In order
to calculate properly the expected number of background counts within our
source extraction ellipse, we extracted total exposure times from both the
source and background regions (with point sources removed) and normalized the
extracted background counts to the source exposure times.  That is, using the
number of background counts $b_{\rm ann}$ and total background exposure time
$T_{\rm ann}$ as measured from the elliptical annulus, we calculated the
expected number of background counts $b_{\rm ext}$ in a source extraction
region with total exposure time $T_{\rm ext}$ as being $b_{\rm ext} = b_{\rm
ann} T_{\rm ext}/T_{\rm ann}$.  This technique therefore accounts for gradients
in the effective exposure over the spatial extent of the extended \xray\ source
and the extracted background region.  We extracted $s_{\rm ext} = 491$~counts
with $b_{\rm ext} = 233$ background counts expected, implying a signal-to-noise
ratio of 11.6$\sigma$.  We then computed the net number of source counts
$n_{\rm src}$ from \hbox{J221744.6+001738} as $n_{\rm src} = (s_{\rm ext} -
b_{\rm ext}) \times A_{\rm src}/A_{\rm ext}$, where the term $A_{\rm
src}/A_{\rm ext}$ is the ratio of the area used to extract counts $A_{\rm ext}$
(i.e., with point-sources masked out) and the total area of the elliptical
extraction region $A_{\rm src}$.  

%
%
\begin{figure}
\centerline{
\includegraphics[width=8.0cm]{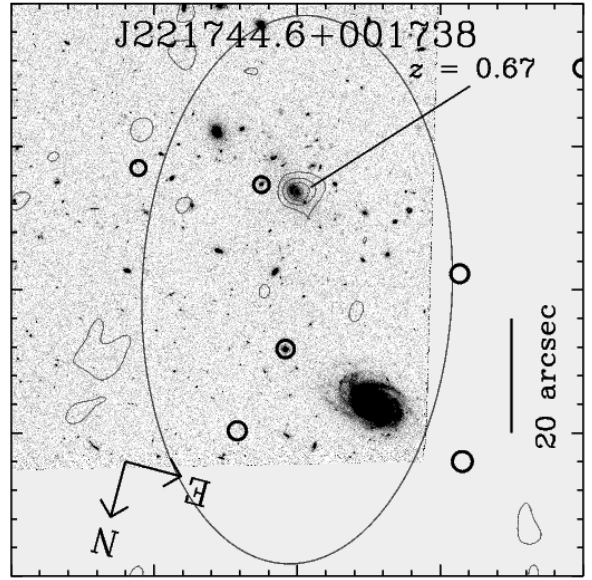}
}
\caption{
\hst\ F814W image of the region within the extended \xray\ source
\hbox{J221744.6+001738} (see $\S$5 for detail).  The image itself is
100~arcsec per side.  The extent of the \xray\ source has been highlighted
with the black ellipse, and represents the approximate shape of the extended
emission that exceeds $\approx$10~per~cent of the local \hbox{0.5--2~keV}
background (determined from the adaptively-smoothed soft-band image; see
$\S$3.1).  \xray\ sources from our main catalog have been highlighted with
thick circles, with the size of each circle representing roughly the 90~per~cent
encircled energy fraction as computed at 1.497~keV.  Within the image, a small
clustering of galaxies is notable in the southern region ({\it top region\/})
of the extended \xray\ emission, which is the likely source of the \xray\
emission.  Radio (1.4~GHz) contours indicating intensity levels of 2.7, 8.0,
13.3, 18.6, 23.9, and 29.2~$\times 10^6$~Jy~sr$^{-1}$ have been indicated.  The
most probable cluster central galaxy \hbox{J221744.3+001722}, which is
coincident with a bright radio source (1.4~GHz flux of 1.4~mJy) and has a
spectroscopic redshift of $z = 0.67$, has been highlighted.
}
\end{figure}

Using the above methods, we find $n_{\rm src} = 410 \pm 37$ soft-band counts;
using the average exposure over the extended source $\approx$363~ks, implies a
soft-band count-rate of $(1.1 \pm 0.1) \times 10^{-3}$~counts~s$^{-1}$.
Figure~13 shows the \hst\ F814W image in the vicinity of the extended \xray\
emission.  From the image, a small clustering of galaxies is apparent in the
southern region of the extraction ellipse, with a few elliptical galaxies
residing in the highest density regions.  It is therefore plausible that the
extended \xray\ emission observed in \hbox{J221744.6+001738} is associated with
a group or poor cluster traced by these galaxies.  We searched the redshift
catalogs discussed in $\S$3.2.3 for obvious redshift spikes; however, due to
small number statistics, we were unable to confirm or deny the presence of an
overdensity in this region.  Using $J-K$ colours, we found that a larger
fraction of the galaxies within the extended \xray\ emission had ``red''
[$(J-K)_{\rm Vega}>1.5$] near-IR colours compared with those found over the
entire \chandra\ observed SSA22 field.  A K-S test indicates that the
distribution of $J-K$ colours in this region are similar only at the
$\approx$6~per~cent confidence level.  The difference in colour distributions
is almost certainly due to an excess of ``red'' galaxies in this region, which
indicate \hbox{$z \approx$~0.5--1.5} galaxies having notable 4000~\AA\ breaks
(e.g., Swinbank \etal\ 2007).  The most likely cluster central galaxy in this
region [selected based on visual morphology, location relative to other nearby
sources, $K$-band magnitude, and near-IR colour; $(J-K)_{\rm Vega} = 2.0$] is
\hbox{J221744.3+001722} (indicated in Fig.~13).   This source has been found to
have a redshift of $z = 0.67$ (from Chapman \etal\ 2009, in-preparation) and is
one of only three sources within the extent of the \xray\ emission to have a
spectroscopic redshift available.  The source is coincident with a bright
1.4~GHz source with a flux density of 1.4~mJy (see contours in Fig.~13; Chapman
\etal\ 2004), which corresponds to a 1.4~GHz luminosity of
$\approx$$3\times10^{24}$~W~Hz$^{-1}$.  Since moderately luminous radio sources
like \hbox{J221744.3+001722} typically trace highly clustered regions (e.g.,
Wake \etal\ 2008), this provides further evidence that the extended \xray\
emission is likely due to the presence of a group or poor cluster.

Assuming that the extended source is indeed powered by a group or poor cluster
at $z = 0.67$, we computed the expected soft-band flux and luminosity assuming
a Raymond-Smith thermal plasma (Raymond \& Smith 1977) with \hbox{$kT =
1.0$~keV} and $Z = 0.2 Z_{\odot}$.  This gives a soft-band flux of $(9.5 \pm
0.9) \times 10^{-15}$~\flux\ and rest-frame \hbox{0.5--2~keV} luminosity of
$(2.3 \pm 0.2) \times 10^{43}$~\xlum.  However, to confirm the redshift and the
intrinsic \xray\ spectral shape of \hbox{J221744.6+001738} requires a more
complete census of galaxy redshifts in this region and more detailed \xray\
spectral analyses, which are beyond the scope of this paper.

%
\section{Summary}
%

We have presented point-source catalogs and basic analyses of sources detected in a
deep $\approx$400~ks \chandra\ exposure over a $\approx$330~arcmin$^2$
region centred on a $z = 3.09$ protocluster in the SSA22 region: The \chandra\
Deep Protocluster Survey.  The survey reaches on-axis flux limits in the
\hbox{0.5--2~keV} and \hbox{2--8~keV} bandpasses of \hbox{$\approx$$5.7\times
10^{-17}$~\xlum} and \hbox{$\approx$$3.0\times 10^{-16}$~\xlum}, respectively.
We have presented a main \chandra\ catalog of 297 point sources, which was
generated by (1) running {\ttfamily wavdetect} at a false-positive probability
threshold of $10^{-5}$ and (2) filtering this list to include only sources that
were determined to have \xray\ emission that was significant in comparison to
their local backgrounds.  In addition to the point sources, we have presented
the properties of one \hbox{0.5--2~keV} extended source, which is likely
associated with a group or poor cluster between \hbox{$z \approx$~0.5--1}.  We
have cross-correlated our main catalog source positions with
near--to--mid-infrared photometry and optical spectroscopic redshift catalogs
to determine the nature of the detected point sources.  The combined \xray\ and
multiwavelength data sets indicate a variety of source types, most of which are
absorbed AGNs that dominate at lower \xray\ fluxes.  In total, we have
determined that 12 of the main catalog sources are likely associated with the
$z = 3.09$ protocluster, including sources that have either spectroscopic
redshifts between \hbox{$z =$~3.06--3.12} and/or sources that are coincident
with LAEs.  The majority of these sources lie in the highest density regions of
the protocluster, and we find evidence that the AGN fraction is positively
correlated with local LAE density (96~per~cent confidence).

\section*{Acknowledgements}

We thank the anonymous referee for their thorough review of the manuscript,
which has improved the quality of the catalogs and this paper.  We gratefully
acknowledge financial support from the Science and Technology Facilities
Council (B.D.L., J.E.G.), the Royal Society (D.M.A., I.S.), and the Leverhume
Trust (D.M.A., J.R.M.).

%
{}
%

\appendix

%
\section{Additional {\ttfamily wavdetect} Sources}
%

In this section, we present the source properties of the additional nine
sources excluded from our main \chandra\ catalog presented in $\S$3.2.2 and
Table~2 that were detected by {\ttfamily wavdetect} at a minimum false-positive
probability threshold $\le$$10^{-6}$.  In Table~A1, we present the properties
of these nine sources.  Table~A1 has the exact same column structure as
Table~2.  Columns have the same meaning as they did in Table~2, with one key
exception: the counts and errors for \xray\ detected sources that are provided
in columns~(8)--(16) come from {\ttfamily wavdetect} photometry.  A source is
considered to be detected in a given band when the source is detected by
{\ttfamily wavdetect}.  When a source is not detected in a given band, an upper
limit is computed in the same manner as that presented in the description of
columns~(8)--(16) in Table~2.  Columns that make use of columns~(8)--(16)
(i.e., column~2 and columns~\hbox{35--43}) utilise the values presented here.

These sources are presented at the convenience of the user who wishes to create
a purely {\ttfamily wavdetect} based catalog of \xray\ sources.  This can be
achieved by appending Table~A1 to Table~2 and filtering the merged catalog by
the desired minimum {\ttfamily wavdetect} probability presented in column~5 to
create a {\ttfamily wavdetect} catalog with false-positive probability
threshold of $10^{-6}$, $10^{-7}$, or $10^{-8}$.  

%
%

\begin{table*}
\begin{minipage}{175mm}
\begin{center}
\caption{Supplementary catalog of additional {\ttfamily wavdetect} sources with false-positive probability $\le$$10^{-6}$ that are not in the main \chandra\ catalog.}
\begin{tabular}{cccccccccccc}
\hline\hline
 & \multicolumn{2}{c}{X-ray Coordinates}  & \multicolumn{2}{c}{Detection Probability} &  & & \multicolumn{3}{c}{Net Counts} \\
Source & \multicolumn{2}{c}{\rule{1.2in}{0.01in}} & \multicolumn{2}{c}{\rule{1.2in}{0.01in}} & Pos. Error & $\theta$ & \multicolumn{3}{c}{\rule{1.8in}{0.01in}}  \\
 Number & $\alpha_{\rm J2000}$ & $\delta_{\rm J2000}$  & {\ttfamily AE} & {\ttfamily wavdetect} & (arcsec) & (arcmin) & 0.5--8~keV & 0.5--2~keV & 2--8~keV  \\
 (1) & (2) & (3) & (4) & (5) & (6) & (7) & (8)--(10) & (11)--(13) & (14)--(16) \\
\hline
   1 &     22 17 02.02 &     +00 13 32.4 &  0.953 & $-$6 &  3.31 &  8.95 &                    7.0$\pm$4.2 &                        $<$19.7 &                        $<$34.8\\
   2 &     22 17 26.23 &     +00 19 34.7 &  0.947 & $-$7 &  1.61 &  4.82 &                        $<$14.3 &                    4.1$\pm$2.2 &                        $<$11.4\\
   3 &     22 17 26.93 &     +00 10 45.2 &  0.802 & $-$6 &  1.82 &  5.41 &                        $<$14.6 &                    4.1$\pm$2.2 &                        $<$11.9\\
   4 &     22 17 28.59 &     +00 18 25.8 &  0.914 & $-$8 &  1.13 &  3.53 &                    4.9$\pm$2.5 &                         $<$7.8 &                    3.7$\pm$2.2\\
   5 &     22 17 31.02 &     +00 07 35.2 &  0.542 & $-$6 &  3.01 &  8.10 &                        $<$27.0 &                    5.4$\pm$3.0 &                        $<$24.4\\
   6 &     22 17 36.58 &     +00 09 11.8 &  0.977 & $-$6 &  1.59 &  6.36 &                   11.0$\pm$4.2 &                         $<$9.9 &                   11.1$\pm$4.1\\
   7 &     22 17 49.43 &     +00 22 18.2 &  0.911 & $-$6 &  2.69 &  7.43 &                    4.9$\pm$4.0 &                        $<$18.0 &                        $<$24.7\\
   8 &     22 18 01.13 &     +00 24 57.7 &  0.660 & $-$6 &  5.45 & 11.20 &                    7.0$\pm$3.9 &                        $<$19.5 &                   15.9$\pm$6.4\\
   9 &     22 18 08.90 &     +00 08 40.5 &  0.887 & $-$6 &  2.49 & 10.57 &                  42.8$\pm$10.4 &                        $<$23.3 &                        $<$39.5\\
\hline
\end{tabular}
\end{center}
NOTE. --- Table~A1 is presented in its entirety in the electronic version; an abbreviated version of the table is shown here for guidance as to its form and content.  The full table contains 44 columns of information for the nine sources.  Meanings and units for all columns have been summarized in detail in $\S\S$~3.2.2 and Appendix~A.
\end{minipage}
\end{table*}

\label{lastpage}


\begin{thebibliography}{}
%

\bibitem[Alexander et al.(2001)]{2001AJ....122.2156A} Alexander, D.~M., 
Brandt, W.~N., Hornschemeier, A.~E., Garmire, G.~P., Schneider, D.~P., 
Bauer, F.~E., \& Griffiths, R.~E.\ 2001, \aj, 122, 2156

\bibitem[Alexander et al.(2003)]{2003AJ....126..539A} Alexander, D.~M., et
al.\ 2003, \aj, 126, 539

\bibitem[Baganoff et al.(2003)]{2003ApJ...591..891B} Baganoff, F.~K., et 
al.\ 2003, \apj, 591, 891 

\bibitem[Boller et al.(1998)]{} Boller, Th., Bertoldi, F., Dennefeld, M., \&
Voges, W.  1998, A\&AS, 129, 87

\bibitem[Bower et al.(2004)]{2004MNRAS.351...63B} Bower, R.~G., et al.\
2004, \mnras, 351, 63

\bibitem[Brandt \& Hasinger(2005)]{2005ARA&A..43..827B} Brandt, W.~N.,
\& Hasinger, G.\ 2005, \araa, 43, 827

\bibitem[Broos et al.(2000)]{} Broos, P., et~al. 2000, User's Guide for the
{\ttfamily TARA} Package.  (University Park: Pennsylvania State Univ.)

\bibitem[Broos et al.(2002)]{} Broos, P. S., Townsley, L. K., Getman, K., \&
Bauer, F. E. 2002, ACIS Extract, An ACIS Point Source Extraction Package
(University Park: Pennsylvania State Univ.)

\bibitem[Chapman et al.(2001)]{2001ApJ...548L..17C} Chapman, S.~C., Lewis, 
G.~F., Scott, D., Richards, E., Borys, C., Steidel, C.~C., Adelberger, 
K.~L., \& Shapley, A.~E.\ 2001, \apjl, 548, L17

\bibitem[Chapman et al.(2004)]{2004ApJ...606...85C} Chapman, S.~C., Scott, 
D., Windhorst, R.~A., Frayer, D.~T., Borys, C., Lewis, G.~F., 
\& Ivison, R.~J.\ 2004, \apj, 606, 85 
 
\bibitem[De Lucia et al.(2006)]{2006MNRAS.366..499D} De Lucia, G., Springel,
V., White, S.~D.~M., Croton, D., \& Kauffmann, G.\ 2006, \mnras, 366, 499

\bibitem[Donley et al.(2008)]{2008ApJ...687..111D} Donley, J.~L., Rieke, 
G.~H., P{\'e}rez-Gonz{\'a}lez, P.~G., \& Barro, G.\ 2008, \apj, 687, 111 

\bibitem[Feigelson et al.(2000)]{} Feigelson, E.D., Broos, P.S., \& Gaffney, J.
2000, Memo on the Optimal Extraction Radius for ACIS Point Sources.  The
Pennsylvania State University, University Park

\bibitem[Freeman et al.(2002)]{} Freeman, P.E., Kashyap, V., Rosner, R., \&
Lamb, D.Q. 2002, ApJS, 138, 185

\bibitem[Garilli et al.(2008)]{2008A&A...486..683G} Garilli, B., et al.\ 2008,
\aap, 486, 683 

\bibitem[Garmire et al. (2003)]{} Garmire, G.~P., Bautz, M.~W., Ford, P.~G.,
Nousek, J.~A., \& Ricker, G.~R.  2003, Proc. SPIE, 4851, 28

\bibitem[Geach et al.(2005)]{2005MNRAS.363.1398G} Geach, J.~E., et al.\
2005, \mnras, 363, 1398

\bibitem[Geach et al.(2009)]{2009arXiv0904.0452G} Geach, J.~E., et al.\ 
2009, \apj, in-press (astro-ph/0904.0452)

\bibitem[Gebhardt et al.(2000)]{2000ApJ...539L..13G} Gebhardt, K., et al.\ 
2000, \apjl, 539, L13 

\bibitem[Gehrels(1986)]{1986ApJ...303...336}
Gehrels, N. 1986, ApJ, 303, 336

\bibitem[Gendreau et al. (1995)]{}Gendreau, K.~C.~et al.\ 1995, \pasj, 47, L5

\bibitem[Georgakakis et al.(2008)]{2008MNRAS.388.1205G} Georgakakis, A., 
Nandra, K., Laird, E.~S., Aird, J., \& Trichas, M.\ 2008, \mnras, 388, 1205 

\bibitem[Getman et al.(2005)]{2005ApJS..160..319G} Getman, K.~V., et al.\ 
2005, \apjs, 160, 319 

\bibitem[Governato et al.(1998)]{1998Natur.392..359G} Governato, F., Baugh,
C.~M., Frenk, C.~S., Cole, S., Lacey, C.~G., Quinn, T., \& Stadel, J.\ 1998,
\nat, 392, 359 

\bibitem[Hayashino et al.(2004)]{2004AJ....128.2073H} Hayashino, T., et
al.\ 2004, \aj, 128, 2073

\bibitem[Hickox \& Markevitch(2006)]{2006ApJ...645...95H} Hickox, R.~C., \&
Markevitch, M.\ 2006, \apj, 645, 95

\bibitem[Hornschemeier et al. (2001)]{} Hornschemeier, A.E., et~al.\ 2001, ApJ,
554, 742

\bibitem[Hornschemeier et al.(2004)]{2004ApJ...600L.147H} Hornschemeier, 
A.~E., et al.\ 2004, \apjl, 600, L147

\bibitem[Jerius et al. (2000)]{} Jerius, D., Donnelly, R.H., Tibbetts, M.S.,
Edgar, R.J., Gaetz, T.J., Schwartz, D.A., Van~Speybroeck, L.P., \& Zhao, P.
2000, Proc. SPIE, 4012, 17

\bibitem[Kauffmann(1996)]{1996MNRAS.281..487K} Kauffmann, G.\ 1996, \mnras,
281, 487

\bibitem[Kim et al.(2006)]{2006ApJ...644..829K} Kim, D.-W., et al.\ 2006, 
\apj, 644, 829

\bibitem[Kim et al.(2007)]{2007ApJS..169..401K} Kim, M., et al.\ 2007, 
\apjs, 169, 401

\bibitem[Lacy et al.(2004)]{2004ApJS..154..166L} Lacy, M., et al.\ 2004, 
\apjs, 154, 166 

\bibitem[Laird et al.(2009)]{2009ApJS..180..102L} Laird, E.~S., et al.\ 
2009, \apjs, 180, 102 

\bibitem[Lawrence et al.(2007)]{2007MNRAS.379.1599L} Lawrence, A., et al.\ 
2007, \mnras, 379, 1599 

\bibitem[Le F{\`e}vre et al.(2005)]{2005A&A...439..845L} Le F{\`e}vre, O., et
al.\ 2005, \aap, 439, 845 

\bibitem[Lehmer et al.(2005)]{2005ApJS..161...21L} Lehmer, B.~D., et al.\
2005, \apjs, 161, 21

\bibitem[Lehmer et al.(2006)]{2006AJ....131.2394L} Lehmer, B.~D., et al.\
2006, \aj, 131, 2394

\bibitem[Lehmer et al.(2007)]{2007ApJ...657..681L} Lehmer, B.~D., et al.\ 
2007, \apj, 657, 681 

\bibitem[Lehmer et al.(2008)]{2008} Lehmer, B.~D., et al.\ 2008, \apj, 681,
1163

\bibitem[Lehmer et al.(2009)]{2009ApJ...691..687L} Lehmer, B.~D., et al.\ 
2009, \apj, 691, 687 

\bibitem[Luo et al.(2008)]{2008ApJS..179...19L} Luo, B., et al.\ 2008, 
\apjs, 179, 19 

\bibitem[Markevitch et al. (2001)]{} Markevitch, M. 2001, CXC memo
(http://asc.harvard.edu/cal/)

\bibitem[Markevitch et al. (2003)]{} Markevitch, M., et~al.\ 2003, \apj, 583,
70

\bibitem[Marshall et al. (1980)]{} Marshall, F.~E., Boldt, E.~A., Holt, S.~S.,
Miller, R.~B., Mushotzky, R.~F., Rose, L.~A., Rothschild, R.~E., \&
Serlemitsos, P.~J.\ 1980, \apj, 235, 4

\bibitem[Matsuda et al.(2004)]{2004AJ....128..569M} Matsuda, Y., et al.\ 
2004, \aj, 128, 569 

\bibitem[Matsuda et al.(2005)]{2005ApJ...634L.125M} Matsuda, Y., et al.\
2005, \apjl, 634, L125

\bibitem[Nandra et al.(2005)]{2005MNRAS.356..568N} Nandra, K., et al.\ 
2005, \mnras, 356, 568 

\bibitem[Ptak et al.(2007)]{2007ApJ...667..826P} Ptak, A., Mobasher, B., 
Hornschemeier, A., Bauer, F., \& Norman, C.\ 2007, \apj, 667, 826 

\bibitem[Raymond \& Smith(1977)]{1977ApJS...35..419R} Raymond, J.~C., \&
Smith, B.~W.\ 1977, \apjs, 35, 419

\bibitem[Richards et al.(2006)]{2006ApJS..166..470R} Richards, G.~T., et 
al.\ 2006, \apjs, 166, 470 

\bibitem[Spergel et al.(2003)]{2003ApJS..148..175S} Spergel, D.~N., et al.\ 
2003, \apjs, 148, 175

\bibitem[Stark et al. (1992)]{} Stark, A.~A., Gammie, C.~F., Wilson, R.~W.,
Bally, J., Linke, R.~A., Heiles, C., \& Hurwitz, M.\ 1992, \apjs, 79, 77

\bibitem[Steidel et al.(1998)]{1998ApJ...492..428S} Steidel, C.~C.,
Adelberger, K.~L., Dickinson, M., Giavalisco, M., Pettini, M.,
\& Kellogg, M.\ 1998, \apj, 492, 428

\bibitem[Steidel et al.(2000)]{2000ApJ...532..170S} Steidel, C.~C., Adelberger,
K.~L., Shapley, A.~E., Pettini, M., Dickinson, M., \& Giavalisco, M.\ 2000,
\apj, 532, 170 

\bibitem[Steidel et al.(2003)]{2003ApJ...592..728S} Steidel, C.~C., Adelberger,
K.~L., Shapley, A.~E., Pettini, M., Dickinson, M., \& Giavalisco, M.\ 2003,
\apj, 592, 728

\bibitem[Swinbank et al.(2007)]{2007MNRAS.379.1343S} Swinbank, A.~M., et 
al.\ 2007, \mnras, 379, 1343 

\bibitem[Tamura et al.(2009)]{2009Natur.459...61T} Tamura, Y., et al.\ 
2009, \nat, 459, 61

\bibitem[Townsley et al.(2000)]{} Townsley, L.K., Broos, P.S., Garmire, G.P.,
\& Nousek, J.A.  2000, ApJ, 534, L139

\bibitem[Townsley et al.(2002)]{} Townsley, L.K., Broos, P.S., Nousek, J.A., \&
Garmire, G.P.  2002, Nuclear Instruments and Methods in Physics Research A,
486, 751

\bibitem[Tozzi et al.(2006)]{2006A&A...451..457T} Tozzi, P., et al.\ 2006,
\aap, 451, 457

\bibitem[Uchimoto et al.(2008)]{2008PASJ...60..683U} Uchimoto, Y.~K., et 
al.\ 2008, \pasj, 60, 683

\bibitem[Vikhlinin et al.(1995)]{1995ApJ...451..553V} Vikhlinin, A., 
Forman, W., Jones, C., \& Murray, S.\ 1995, \apj, 451, 553 

\bibitem[Vikhlinin et al. (2001)]{} Vikhlinin, A. 2001, CXC Memo, (Cambridge:
CXC) http://asc.harvard.edu/cal/Acis/Cal\_prods/vfbkgrnd/

\bibitem[Wake et al.(2008)]{2008MNRAS.391.1674W} Wake, D.~A., Croom, S.~M., 
Sadler, E.~M., \& Johnston, H.~M.\ 2008, \mnras, 391, 1674 

\bibitem[Webb et al.(2009)]{2009ApJ...692.1561W} Webb, T.~M.~A., Yamada, 
T., Huang, J.-S., Ashby, M.~L.~N., Matsuda, Y., Egami, E., Gonzalez, M., 
\& Hayashimo, T.\ 2009, \apj, 692, 1561 

\bibitem[Weisskopf et al.(2007)]{2007ApJ...657.1026W} Weisskopf, M.~C., Wu, 
K., Trimble, V., O'Dell, S.~L., Elsner, R.~F., Zavlin, V.~E., 
\& Kouveliotou, C.\ 2007, \apj, 657, 1026

\bibitem[Wilman et al.(2005)]{2005Natur.436..227W} Wilman, R.~J., Gerssen,
J., Bower, R.~G., Morris, S.~L., Bacon, R., de Zeeuw, P.~T.,
\& Davies, R.~L.\ 2005, \nat, 436, 227

\bibitem[Worsley et al.(2005)]{2005MNRAS.357.1281W} Worsley, M.~A., et al.\ 
2005, \mnras, 357, 1281 

%
\end{thebibliography}
\end{document}